\documentclass[10pt,twocolumn]{article}
\usepackage{simpleConference}
\usepackage{times}
\usepackage{graphicx}
\usepackage{amssymb}
\usepackage{stfloats}
\usepackage{todonotes}
\usepackage{float}
\usepackage{enumitem}
\definecolor{ForestGreen}{RGB}{34,139,34}

\usepackage{xurl}
\usepackage{caption}
\DeclareCaptionFont{8pt}{\fontsize{8pt}{9pt}\selectfont}
\usepackage[normalem]{ulem}

\usepackage{hyperref}
\hypersetup{
           breaklinks=true,
           colorlinks=true
}

\title{A case study on latency, bandwidth and energy efficiency of mobile 5G and YouTube Edge service in London. \\Why the 5G ecosystem and energy efficiency matter?}

\author{Peixuan Song \\
University of Cambridge\thanks{Author's study conducted during the final year of his undergraduate degree in 2022.} \\
ps828@cantab.ac.uk \\
\and
JunKyu Lee \\
University of Essex\\
j.lee@essex.ac.uk \\
\and
Lev Mukhanov \\
Queen Mary University of London\\
l.mukhanov@qmul.ac.uk  \\
}
\date{}

\begin{document}

\maketitle
\begin{abstract}
The advancements in 5G mobile networks and Edge computing offer great potential for services like augmented reality and Cloud gaming, thanks to their low latency and high bandwidth capabilities. 
However, the practical limitations of achieving optimal latency on real applications remain uncertain.
This paper aims to investigate the actual latency and bandwidth provided by 5G Networks and YouTube Edge service in London, UK. We analyze how latency and bandwidth differ between 4G LTE and 5G networks and how the location of YouTube Edge servers impacts these metrics.  Our research reveals over 10 significant observations and implications, indicating that the primary constraints on 4G LTE and 5G capabilities are the ecosystem and energy efficiency of mobile devices down-streaming data. Our study demonstrates that to fully unlock the potential of 5G and it's applications, it is crucial to prioritize efforts aimed at improving the ecosystem and enhancing the energy efficiency.
\end{abstract}

\section{Introduction}
It is projected that the number of intelligent Internet-connected devices will soon reach tens of billions~\cite{iot_devices}. All these devices use Internet and Cloud data centers to transfer and store the data. As a result, the size of data transferred through Internet will exceed 24.3 exabytes soon~\cite{cisco_report}. Today, mobile devices are one of the primary source of the data transferred to and from Cloud; the mobile internet traffic, which is generated by 4.7 billion mobile Internet users worldwide, is about 57\% of the global online traffic~\cite{global_traffic}. 
The performance of numerous mobile applications, including YouTube, navigation services, and games, is highly reliant on network latency and bandwidth, which should be sufficient to meet user expectations effectively. Moreover, a new class of emerging applications, such as interactive augmented reality and Cloud gaming applications, generate much more data than the state-of-the-art applications and are more critical to the network latency and bandwidth. 

Edge computing was introduced to address the challenge of increasing mobile data traffic~\cite{ibm_report}. By bringing computing resources closer to the edge of the network, Edge computing can help to reduce latency and increase bandwidth, and, thus, improve the performance of applications that require real-time processing. This can be especially beneficial for applications such as augmented reality and self-driving cars.
Google has already implemented 7500 Edge servers to enable the Stadia service, a Cloud gaming service~\cite{8818578}, which however was deprecated recently. In our study, we also demonstrate that Google also places Youtube Edge servers nearby base stations. However, the actual latency and bandwidth that can be achieved in real applications that utilize 5G networks and Edge servers in UK remain unclear. While some studies suggest that 5G networks can theoretically achieve an average latency as low as 1 ms~\cite{5g_latency}, real-world implementation may encounter significant overhead due to the 4G/5G ecosystem, i.e. base stations, wired/fiber communication network and Cloud/Edge servers.

The \textbf{main goal} of our study is to investigate what is the maximum latency and bandwidth can be achieved by real mobile applications that use 5G networks and Edge servers in London. We want to understand how location of servers handling data requests from mobile devices affect the latency. We also aim to investigate how down-streaming affects the energy consumption of mobile devices. By studying all these aspects, we want to understand what emerging 5G applications can be enabled nowadays in London and  what factors prevent exploiting full capabilities of 4G LTE and 5G networks.

To enable our study, we use a Google Pixel 4a smartphone which supports 5G networks. We test the latency and bandwidth provided the 5G and 4G LTE networks when down-streaming YouTube videos for three major mobile network operators. We track the location of servers which handle data requests and measure down-streaming latency and bandwidth, as well as the energy consumed by the mobile device. We use the collected data to analyze the relationships between latency, bandwidth and geo-graphical locations of servers. In addition, we measure the device current and power to investigate energy efficiency of the mobile device when down-streaming YouTube videos. Given the fact that YouTube is one of the most popular services in the world with billions of users\cite{youtube_ref},  we believe that it is well optimized to provide the best possible availability, latency and bandwidth. Thus, YouTube can serve as a reliable indicator of the expected 5G network quality when running real applications that utilize Edge and Cloud servers. Based on the results of our experiments, we make conclusions about which emerging 5G applications can be implemented in London, UK.

The contribution of this paper can be summarized as follows:
\begin{itemize}[leftmargin=*,noitemsep,topsep=0pt, partopsep=0pt]
    \item We present the results of our study on latency, bandwidth and energy efficiency for down-streaming in 4G LTE and 5G mobile networks operated by three major operators in London, UK. 
    \item We make several important observations about latency and bandwidth, such as: i) the smallest average latency was obtained for 4G LTE, approximately 25 ms, as opposed to 5G; ii) the average 5G latency varies from 37 ms to 150 ms depending on server location and mobile operator; iii) the minimum latency measured in our experiments for both 4G LTE and 5G is 20.7 ms and 24.3 ms, respectively; iv) the average down-streaming bandwidth for 4G LTE and 5G networks typically range around 0.24 Gbps and 0.5 Gbps, respectively.
    \item We reveal that the 4G/5G ecosystem is a major bottleneck which prevents to fully enable capabilities of 4G LTE and 5G networks. By enhancing the ecosystem, the average latency of can be reduced by up to 2$\times$. We demonstrate that enabling 1ms latency for emerging 5G applications will require a network of Edge servers, distance between which does not exceed 227 km.
    \item We highlight that energy efficiency is another major bottleneck preventing implementation of mobile applications that utilize 4G LTE and 5G. To be more specific, we show that the smartphone current and power increases by 68\% on average when down-streaming data and, as a result, some mobile games, such as Realm Grinder, can consume more energy when running on NVIDIA Cloud/Edge servers compared to the version of these games which use only mobile GPUs. Moreover, we demonstrate that enabling global YouTube down-streaming on mobile devices would require significant energy demands, equivalent to a nuclear plant.  
    \item Finally, we discuss which applications can benefit from 5G networks. Our findings indicate that existing London-based 4G LTE and 5G networks, along with the smartphones used in our study, fail to meet the latency requirements for crucial applications such as Autonomous Driving Vehicles and AR/VR 3D rendering.
\end{itemize}

\textbf{\textit{Overall, based on the results of our study, we conclude that fully enabling crucial 5G applications necessitates substantial improvements in the ecosystem and the energy efficiency of 5G modems.}}

The paper is organized as follows: Section \ref{sec:background} presents background and related work; Section \ref{sec:exp_setup} presents our experimental framework and  methodology; Section \ref{sec:experiments} presents the results of out experimental study; Section \ref{sec:discussion} discusses the energy efficiency challenge and if the obtained latency and bandwidth measurements meet the demands of emerging 5G applications; Section \ref{sec:limitation} demonstrates the limitations of our study and Section \ref{sec:concl} presents the conclusion.

\section{Background and Related Work}
\label{sec:background}
In 2022, the exponential growth of Internet-connected mobile devices resulted in the generation of a staggering 129.4 Exabytes of data, necessitating its processing in Cloud centers\cite{global_intetnet}. However, many user mobile services, such as navigation, Cloud gaming and Augumented reality applications, need to receive a reply from servers with a small latency. For example, Cloud gaming implies that the graphical pipeline runs remotely on Cloud servers \cite{10.1145/2742647.2742656}.    

Introducing 5G networks should significantly improve latency and bandwidth for mobile data transfers. However, the data should travel between base stations and Cloud data centers which can negatively affect latency and bandwidth. To address this issue, the concept of Edge computing was introduced which proposes to place servers close to base stations\cite{edge_copmuting}. In our study, we aim to understand what latency and bandwidth can be achieved over 5G networks using Edge servers and real applications in practice today. To this end, we test 5G network characteristics in London, UK, using YouTube mobile application. We specifically use YouTube since this is one of the most popular global service, having 2.56 billion of active users\cite{youtube_ref}, which should be well optimised to achieve best latency and bandwidth characteristics. 

There are several experimental studies which measure the propagation of mmWaves in suburban and vegetated
environments\cite{9987496,electronics9111867,https://doi.org/10.1002/ett.3311,8641429,7811193,7954664,8419179}. These studies comprehensively investigate mmWave propagation under various conditions, encompassing urban environments, suburban and vegetated areas, human body blockage, and rain-induced fading. While the main goal of our study is to estimate latency and bandwidth of 5G that can be achieved in commercial applications in London, UK. 

Previous studies tried to investigate possible characteristics of 5G networks in London based on the location of base stations\cite{8647379}. However, this study uses a simulation-based framework to estimate latency and bandwidth and does not take into the consideration the delays required by Cloud/Edge networks to process the user requests and send the data back. Recent papers present the results of extensive research studies on 5G latency, bandwidth and energy efficiency measured on real devices \cite{10.1145/3387514.3405882,10.1145/3366423.3380169,10.1145/3452296.3472923, 10.1145/3419394.3423629}. However, these studies use custom applications (or \textit{Speedtest}) and servers to test the quality of 5G networks.  Meanwhile, in our study, we aim to investigate the characteristics of the commercial YouTube application and commercial servers which are integrated by mobile operators with base stations. Thus, we are able to estimate latency and bandwidth which can be achieved by commercial applications using Edge and Cloud servers in practice.

\section{Experimental setup}
\label{sec:exp_setup}
In this section, we present our experimental setup.
\begin{table}[t]
\small
\centering
\begin{tabular}{c} \hline
\textbf{Description}\\ \hline
\vspace{0.1cm}
\textbf{\textit{Processor:}} Snapdragon 765G 5G \\
\textbf{\textit{Cores:}} 1x2.4GHz Cortex-A76, 1x2.2GHz Cortex-A76, \\ 6x1.8GHz Cortex-A55\\
\textbf{\textit{L3 cache/DRAM:}} 2.00 MB/6GB LPDDR4X\\
\textbf{\textit{GPU:}} 3x750MHz Adreno 620\\
\textbf{\textit{Network technology:}} GSM / HSPA / LTE \\
and 5G (Sub-6 and mmWave)\\
\textbf{\textit{OS:}} Android 10\\ \hline
\end{tabular}
\caption{Specifications of Google Pixel 4a/5G.}
\label{tab:hardware_characteristics}
\end{table}

\begin{figure}[t]
        \begin{minipage}[t]{0.48\linewidth}
                \includegraphics[width=\columnwidth]{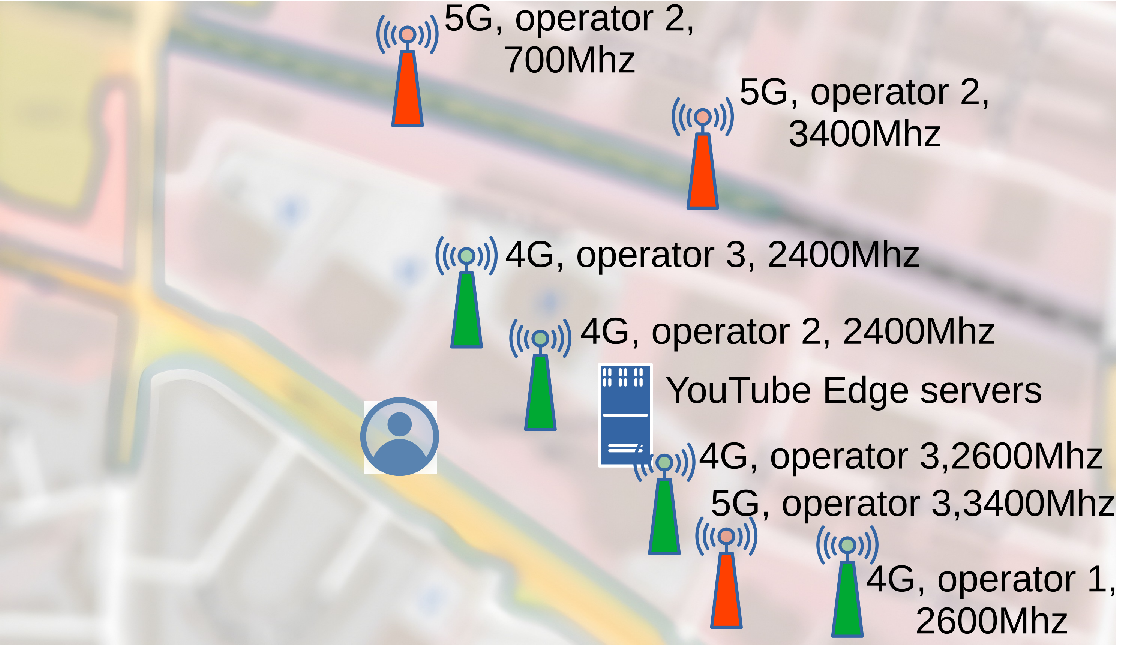}
                \caption{Location of base stations.}
                \label{fig:experiments_map}
        \end{minipage}%
        \hfill%
        \begin{minipage}[t]{0.48\linewidth}
                \includegraphics[width=\columnwidth]{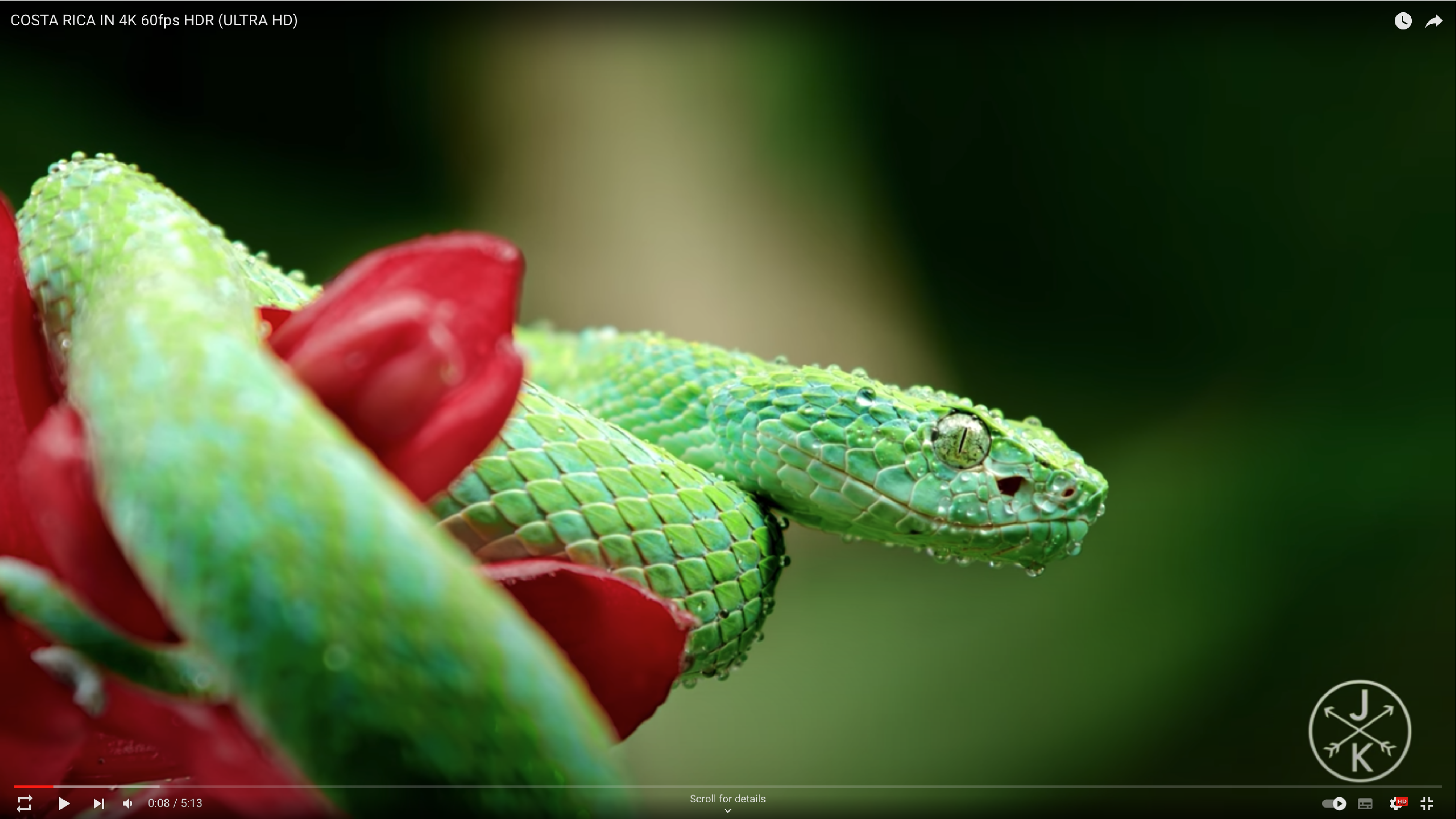}
                \caption{Video screenshot.}
                \label{fig:thermal:b}
        \end{minipage}
\end{figure}

\textbf{The tested mobile device.}
To enable our study, we use a Google Pixel 4a (5G) mobile device which supports 4G LTE (Advanced) and 5G networks. The specification of the device is provided in Table \ref{tab:hardware_characteristics}.

\textbf{YouTube service.}
To test latency of mobile 5G networks, we use mobile YouTube application down-streaming data from Edge servers\cite{youtube_edge}. In particular, we test the mobile network using YouTube application since it is one of the most demanding commercial application to latency and bandwidth. We perform our experiments by down-streaming a 5-minute YouTube video, which is available by the following link \footnote{https://www.youtube.com/watch?v=LXb3EKWsInQ}. We specifically use this video, since it contains contrast figures in the HDR (Ultra HD) format; the maximum resolution is 4096p at 60 FPS(\textit{Frame Per Second}). To test latency under different conditions, we use different resolutions of the video, i.e. 360p ( 480 x 360 pixels), 720p ( 1280 x 720 pixels), 1080p ( 1920 x 1080 pixels) and 4096p ( 4096p x 2160 pixels). In each experiment, we are down-streaming the video 5 times to obtain representative measurements.

\textbf{Latency profiling.} To measure the network latency, we use the Round-Trip Time (\textit{RTT}) metric\cite{10.1145/1698750.1698753}. RTT is the time between a moment when a package is sent to a destination address and a moment when the acknowledgement is received from the destination address.  

To transfer the data over 4G LTE and 5G networks, Android uses \textit{User Datagram Protocol (UDP)} but not \textit{Transmission Control Protocol (TCP)} \cite{6524403}. Android devices also use QUIC (\textit{Quick UDP Internet Connections}) as the encrypted transport layer on the top of UDP, which improves the Quality-Of-Experience (\textit{QOE}) and reduces the latency \cite{10.1145/3098822.3098842}.

In the TCP protocol, RTT can be measured by the time between a packet is sent and its acknowledgment received. However, QUIC is encrypted and multiplexed which makes it hard to measure RTT since a QUIC packet and its acknowledgment might not be on the same path \cite{10.1145/3098822.3098842}. Nonetheless, QUIC accurately calculates the Round-Trip Time (RTT) by including in its acknowledgment packets the time it takes to receive the packet and send the acknowledgment. This helps QUIC estimate the total RTT for different paths, considering both the inbound and outbound paths.
Since QUIC acknowledgments are encrypted and only the RTT from the initial handshake is visible, we use it to estimate latency.

To measure bandwidth and latency, we run \textit{tcpdump} tool \cite{tcpdump} with the following parameters:

\textit{tcpdump -vv -i any -s 0 -w /sdcard/cap.pcap}

To estimate RTT, we parse the log files provided by \textit{tcmpdump}. Meanwhile, to capture bandwidth and latency, we use a specific network protocol analyzer tool,\textit{Wireshark}\cite{5496372}. 

\textbf{Bandwidth profiling.}
To estimate bandwidth, we also parse the QUIC traces and record the number of bytes received every second. Subsequently, we calculate the average bandwidth by aggregating data over one-second intervals. Note that to measure the peak bandwidth, we aggregate data over 1 ms periods and project it for a one-second period.

\textbf{Energy profiling.}
To profile energy consumption, we use \textit{Perfetto}\cite{perfetto}. \textit{Perfetto} uses the data exposed through the charge counters in \textit{Android IHealth HAL} to get the battery current\cite{android_hal}. In particular, this framework measures current in micro ampere, $mA$, within a small period of time. Note that we measure the current for the entire mobile device, which includes SOC (i.e. the processor), screen and 4G LTE/5G modem.
We estimate energy consumption using the current measurements and battery voltage, which is 3.85V on average.

\textbf{Finding the location of the servers.}
To find the location of the servers, we parse the QUIC logs and extract information about the IP addresses down-streaming the video. We identify the location of the servers by IP using six web-services \footnote{https://www.ip2location.com/; https://ipapi.co/; https://ipinfo.io/; https://ipgeolocation.io/; https://ipregistry.co/; https://db-ip.com/}. Unfortunately, IP geo-location can be inaccurate due to dynamic IP addresses, VPNs, shared IP addresses, inaccurate geo-location databases and limited information \cite{10.1145/3404868.3406664,10.1145/3131365.3131380,10.1007/978-3-540-71617-4_26,10.1145/1971162.1971171}. We use the results of geo-location services to filter the measurements taken in our experiments; in particular, we remove all the measurements if the same IP address points to different locations in different services. At the second stage of our filtering, we remove all the experiments with the down-streaming servers for which latency is lower than the time required for the light to travel, in the fiber\footnote{The speed of light in the fiber is $2.18 \times 10^{8}$ m/s}, to the location of these servers and back\cite{10.1145/3131365.3131380,10.1007/978-3-540-71617-4_26,10.1145/1971162.1971171}.    

\textbf{Location and mobile operators.} 
In our experiments, we test 4G LTE (Advanced) and 5G mobile networks for three major mobile operators, each serving approximately 30\% of mobile users in UK. Thus, we expect that the base stations have almost the same load. To the best of our knowledge, all three operators use 5G NSA (\textit{Non-Standalone})\cite{8647379}, which rely on 4G networks to send the control information, at the moment when our study was done, i.e. March 2022.

We test mobile networks in the border area near London. We choose specifically this area since the local London servers which stream the video for 2 operators are located in the same building in this area, according to the geo-location web services (see Figure \ref{fig:experiments_map}). Furthermore, this building also serves as the location for the base stations of these operators (see \textit{operator 2} and \textit{operator 3} in Figure \ref{fig:experiments_map}). We use an online service\footnote{https://www.cellmapper.net/} to find the position of base stations and estimate the signal strength at our specific location. Although the local servers for \textit{operator 1} are located in a different area, one of the base station of this operator is only 200 meters away from the building. Such a location allows us to minimize the distance between the servers, base stations, and mobile devices. Note that \textit{operator 1} does not provide 5G service in this area. At the location where we make our experiments, all three operators provide strong phone signals, ranging from -40 dBm to 85 dBm. All the operators use \textit{low-band} and \textit{mid-band} base stations in this area for 4G and 5G (see Figure 
\ref{fig:experiments_map}). Each \textit{band} operates a different spectrum of wave frequencies which defines the maximum possible bandwidth and minimum latency\cite{band_spectrum,latency_spectrum}. Note that there are many more parameters, such as channel bandwidth, modulation and types of MIMO antennas, that can affect bandwidth and latency\cite{s22010026,malviya_panigrahi_kartikeyan_2017}. Table \ref{tab:band-types} shows spectrum of operating frequencies, the maximum bandwidth and minimum latency for each band \cite{band_spectrum, band_spectrum2, latency_spectrum}. To the best of our knowledge, as of March 2022, there were no publicly available 5G high-band stations in London during the time of our experiments.
\begin{table}[t]
\scriptsize
\centering
\begin{tabular}{|c|c|c|c|c|} 
    \hline 
    \textbf{Bands} & \textbf{Frequency}& \textbf{Support} & \textbf{Bandwidth} & \textbf{Latency}\\
    \hline
     Low-bands & $<$ 1GHz & 5G/4G & 50-100 Mbps & 20ms\\
    \hline
     Mid-bands & 1GHz-6GHz & 5G/4G & 100-900 Mbps & 10ms\\
    \hline
     High-bands & 24GHz-47GHz & 5G & 1-10 Gbps & 1ms\\     
    \hline
\end{tabular}
\caption{Wave frequency spectrum for 5G and 4G.}
\label{tab:band-types}
\end{table}

Importantly, since the experimental area is sparsely populated, we expect that mobile network load will be low. Considering all the facts provided above, we believe that the area where we make our experiments provide optimal conditions for testing maximum performance of commercial mobile networks. 
\section{Experiments}
\label{sec:experiments}
\subsection{IP-based geo-location}
\begin{figure}[htb]
        \centering
        \includegraphics[width=0.7\columnwidth, keepaspectratio]{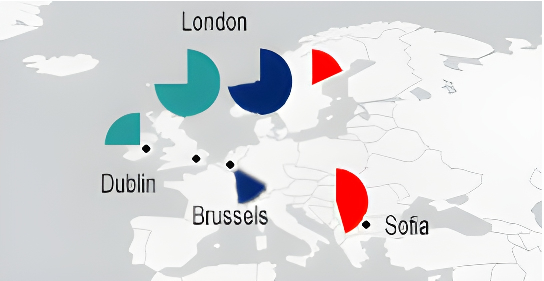}
        \includegraphics[width=0.345\columnwidth, keepaspectratio]{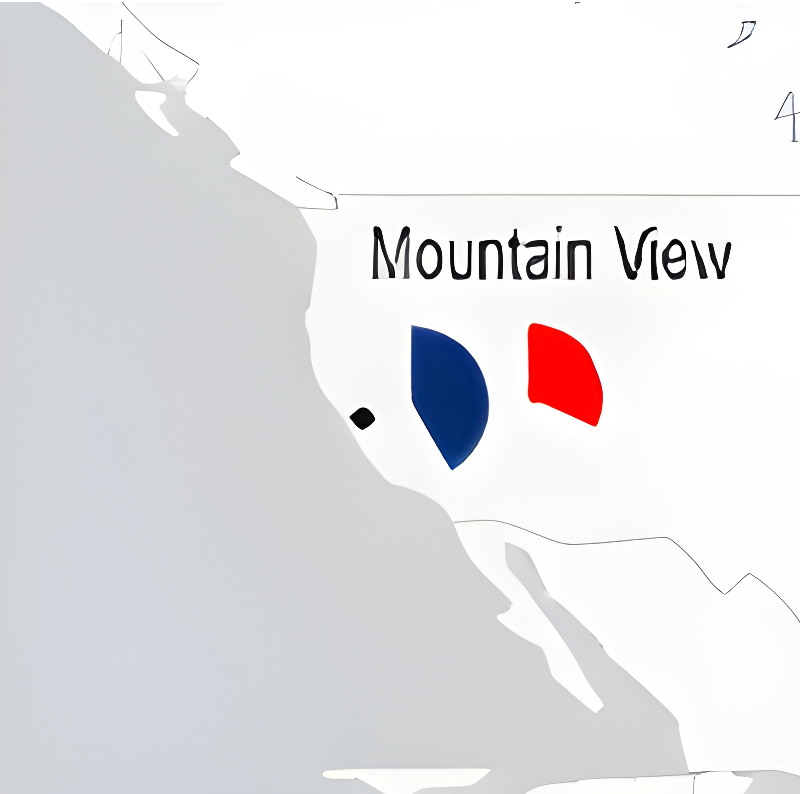} \includegraphics[width=0.345\columnwidth, keepaspectratio]{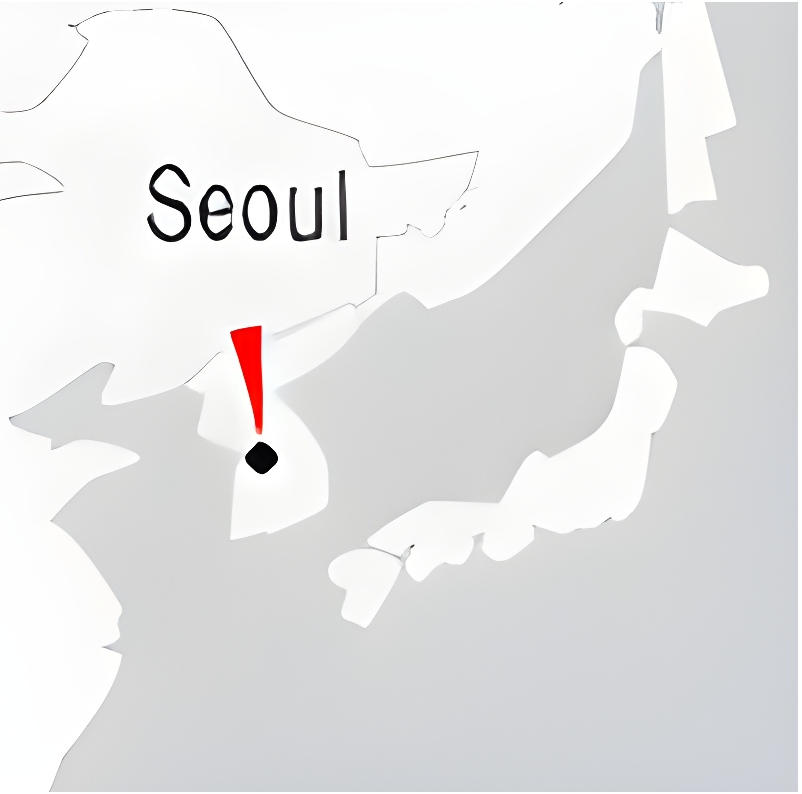}
        \caption{Location of the servers streaming the video.}
        \label{fig:loc}
\end{figure}
We start our study from investigating the location of the servers which stream the video when we run the YouTube application in the area discussed above. To find location of the YouTube servers, we down-stream the video using 4G LTE and 5G and trace the IP addresses of the servers streaming the video by parsing the \textit{tcpdump} logs. Note that a video streamed by YouTube is split into packages which are downloaded by YouTube application with some period. The YouTube service is organized in such a way that each package can be sent from different servers. We identify all the IP addresses of the servers streaming the test video in our experiments. Figure \ref{fig:loc} shows location of YouTube servers streaming the video for three UK mobile operators. Within this figure, each location is depicted with a pie chart indicating the percentage of servers used by a specific operator that were streaming the video from that particular location. The distinct colors within the chart correspond to different mobile operators.

\textbf{Observation1:} \textit{Based on our initial observation, it is evident that the servers streaming the video are not exclusively located in London. Although a significant number of servers are situated in London, we have noticed instances where video streaming has taken place through servers in other locations, such as Sofia, Brussels, and even Mountain View (California).}

The distribution of the streaming servers among different countries for each operator and type of the network is summarized in Table \ref{tab:location}. Our observation reveals that for \textit{operator 1} and \textit{operator 2}, all the servers streaming the video in 4G are located in UK (London) or Europe (Dublin for \textit{operator 1} and Brussels for \textit{operator 2}). Meanwhile, \textit{operator 3} utilize servers mostly located in Europe, Asia and USA. For 5G networks, we also see that the servers from different regions were used. Importantly, all the servers streaming the video belong to Google, according to the used IP geo-location services. 

\textbf{Observation2:} \textit{The location of servers streaming the video to the mobile device depends on mobile operators.}

Interestingly, the number of servers streaming the video varies across operators and network types. As an example, \textit{Operator 2} employes four servers, which are situated in both London and Brussels, to facilitate video streaming via 4G LTE. However, when utilizing 5G technology, the same operator relies on only three servers located in London and Mountain View for video streaming.
Meanwhile, \textit{operator 3} uses 13 and 9 different servers distributed across UK, Europe, Asia and USA for 4G LTE and 5G. We presume that the servers and streaming route are chosen by Google based on the facilities which each operator has and the network load.

\begin{table}[t]
    \scriptsize
 	\centering
\begin{tabular}{ |c|c|c|c|c|c|c| } 
 	 \hline 
 	 \textbf{Operator} & \textbf{Network} & \textbf{\# of IPs} & \textbf{UK} & \textbf{Europe} & \textbf{USA} & \textbf{Asia}\\ 
 		\hline
 	 \#1 & 4G & 4 & 75\% & 25\% & 0\% & 0\% \\
 		\hline
 	 \#2 & 4G & 4 & 75\% & 25\% & 0\% & 0\% \\
 		\hline
 	 \#2 & 5G & 3 & 67\% & 0\% & 33\% & 0\% \\
 		\hline
 	 \#3 & 4G & 13 & 23\% & 38\% & 30\% & 8\% \\
 		\hline
 	 \#3 & 5G & 9 & 11\% & 56\% & 33\% & 0\% \\
 		\hline
\end{tabular}
 	\caption{The distribution of servers streaming the video among different countries.}
	\label{tab:location}
\end{table}

\subsection{Latency}
\begin{figure}[htb]
        \centering
        \includegraphics[width=0.8\columnwidth, keepaspectratio]{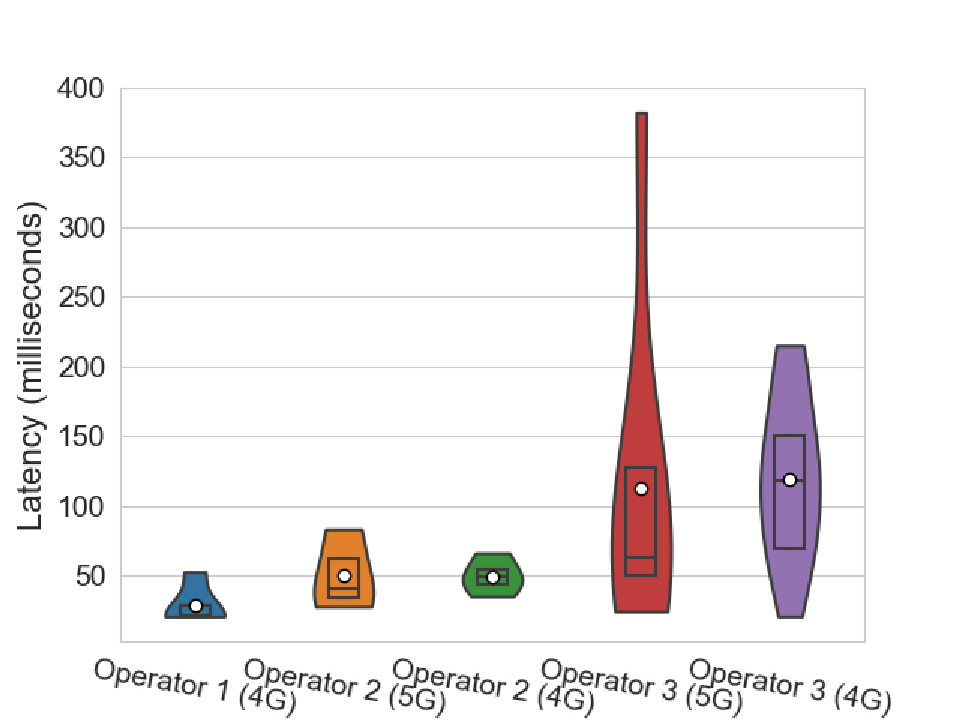}
        \caption{Latency distribution for different operators and types of networks.}
        \label{fig:violin}
\end{figure}

Figure \ref{fig:violin} presents the distribution of measured down-streaming latencies for each operator and network. We present the distribution as a violin plot which depicts the median values (black lines within rectangles), the average values (white dots), the interquartile ranges (black rectangles) and the densities of measured latencies. Surprisingly, we observe the smallest average latency, which is about 25 ms, for the first operator for which only the 4G LTE network is available at the location where we made the measurements; the average 5G latency is about 110 ms and 59 ms for the second operator and and third operator, respectively.

\textbf{Observation3:} \textit{When comparing different operators, the smallest average latency was achieved with 4G LTE, not 5G. This latency was at least half the minimum average latency observed in 5G networks.} 

When comparing average latencies between 4G LTE and 5G networks from the same operator, we observe that in case of \textit{operator 3}, 5G latency is only 10\% lower than that of 4G LTE. However, for \textit{operator 2}, the latencies are identical.

\textbf{Observation4:} \textit{When comparing measurements taken for the same operator, it was observed that the average latency of video down-streaming on 5G is only 10\% lower than the latency experienced on 4G LTE.}

We see that the smallest median latency is also obtained for \textit{operator 2} which provides only 4G LTE service. However, the median latency, which is about 20 ms, is slightly lower than the average latency, i.e. 25 ms. The median latencies coincide with the average latencies for 4G LTE provided by \textit{operator 2} and \textit{operator 3}. These result imply that the down-streaming latencies are evenly distributed around the average values. 
We also see that the highest distribution of the latencies is obtained for both 4G LTE and 5G services provided by \textit{operator 3}. Importantly, this operator uses the highest number of different servers which are located outside UK. According to Table \ref{tab:location}, it appears that only 23\% of the servers used by this operator for 4G LTE and 11\% for 5G are located in the UK. 

\textbf{Observation5:} \textit{The highest distribution of down-streaming latencies is obtained for 4G LTE and 5G networks where the operators use servers distributed across several countries.} 

\begin{figure}[H]
        \centering
        \includegraphics[width=0.8\columnwidth, keepaspectratio]{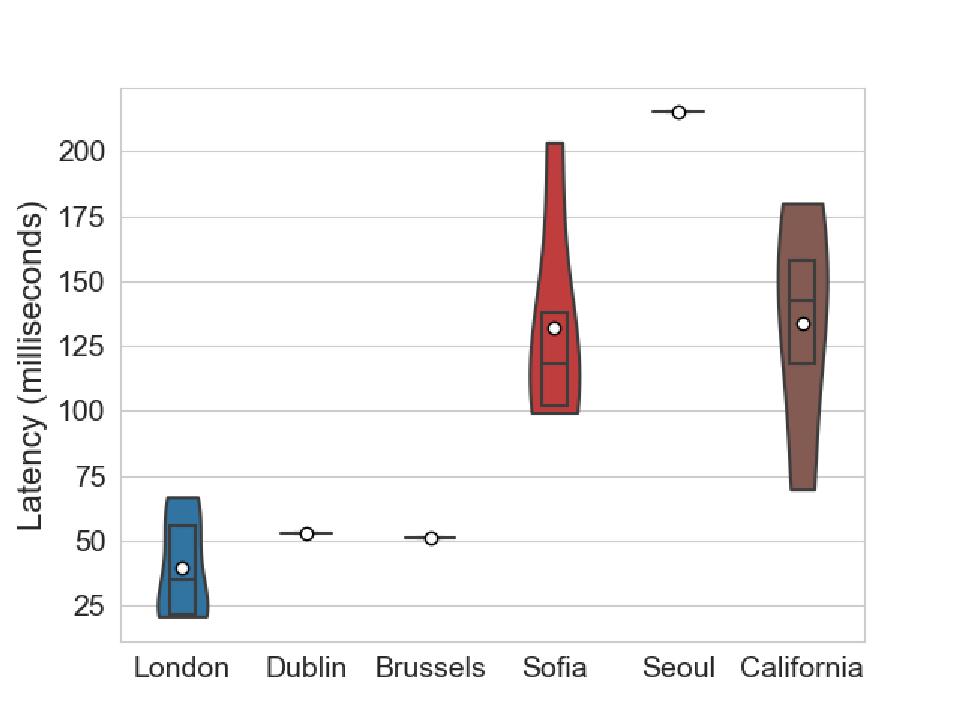}
        \caption{QUIC latency distribution for different locations obtained for 4G LTE.}
        \label{fig:quic-locs-4g}
\end{figure}

To further investigate how latency changes with location of servers, we construct latency distributions for each city where the servers are positioned. Figure \ref{fig:quic-locs-4g} illustrates latency distribution across cities in our experiments, for 4G LTE. As anticipated, the video down-streams originating from London exhibit the lowest average latency, approximately 35 ms. Interestingly, this latency is 1.4$\times$ higher than the average 4G LTE latency for \textit{operator 1}, which uses servers located in London and Dublin. These results imply that, in addition to the location, the major factor which affects latency is the 4G/5G ecosystem, i.e. base stations, internal network hardware and configuration, and Cloud/Edge servers. 

We see that the average latency for servers located in Sofia and Mountain View (California) is about 135 ms, which is $3.85\times$ higher than the average latency obtained in London. We also see that the variation of latencies measured for London servers is lower compared to the variations of the latencies obtained for Sofia and Mountain View. Thus, we may conclude that the use of the servers located in London reduces not only the average latency but also the variation of latencies.

Although the average latency for Dublin and Brussels is nearly twice as high as the average latency in London, the limited number of measurements obtained for these cities makes it difficult to accurately quantify this difference. There are also not enough measurements for Seoul servers with access latency exceeding 200 ms to draw meaningful conclusions.

\begin{figure}[H]
        \centering
        \includegraphics[width=0.8\columnwidth, keepaspectratio]{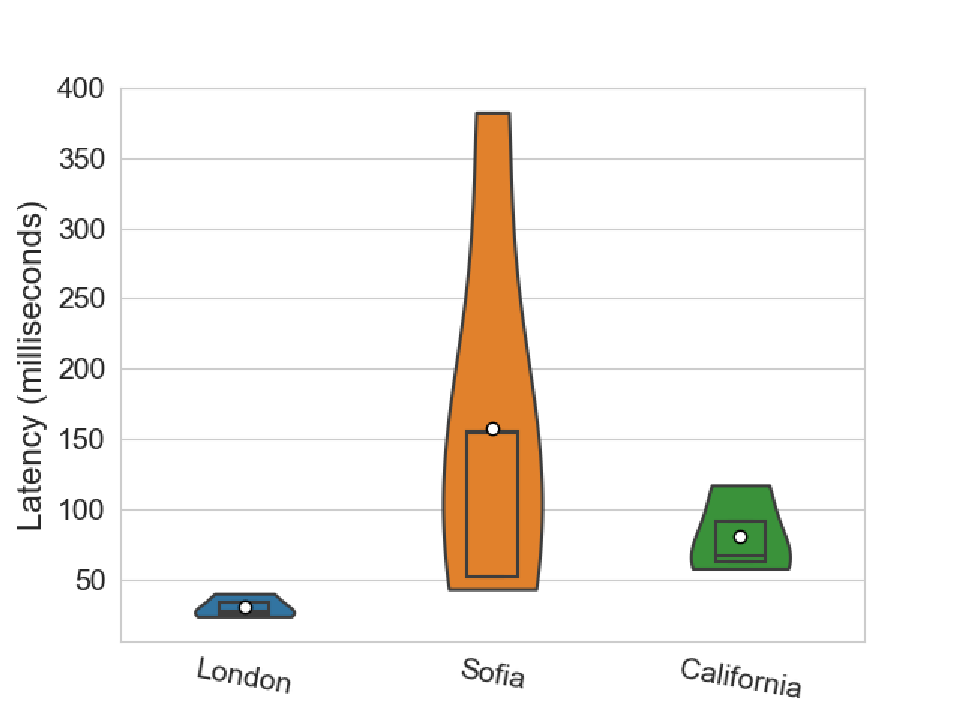}
        \caption{QUIC latency distribution for different locations obtained for 5G.}
        \label{fig:quic-locs-5g}
\end{figure}

Figure \ref{fig:quic-locs-5g} shows the 5G latency variation across different cities. Similar to the experiments with 4G LTE, we observe the lowest average latency, which is about 37 ms, for London. 
Moreover, the 5G average latency, which is 150 ms, is almost the same as the 4G latency measured for the servers located in Sofia. However, the 5G latency is only 75 ms which is almost $2\times$ smaller than than the 4G latency measured for servers located in Mountain View. We attribute this to the superior network infrastructure employed by 5G for connecting servers in London and Mountain View. This also clarifies why the average 5G latency obtained for servers in Mountain View is $2 \times$ smaller than the 5G latency experienced by servers in Sofia. It is important to note that Sofia is only 2000 km away from London, whereas Mountain View is situated 8631 km apart. This implies that the disparity in latencies is also due to the 4G/5G ecosystem. 

We use the average latencies measured for different cities to project the maximum distance between servers and mobile devices to achieve a particular latency. To be more specific, we use the difference in the average latency obtained for London and Mountain View to estimate how the latency increases with distance as follows:
\begin{equation}
Lat\_per\_km = \frac{Lat^{Mountain View}_{Avr} - Lat^{London}_{Avr}}{distance^{London}_{Mountain View}}
\label{eq:lat_per_distance}
\end{equation}
Note that we specifically rely on the 5G measurements taken for Mountain View, as they represent the minimum latency increase per km. Thus, we project the maximum distance required to achieve a specific latency for the best case scenario.

\begin{table}[t]
\scriptsize
\centering
\begin{tabular}{|c|c|} 
    \hline 
    \textbf{Latency} & \textbf{Distance} \\
    \hline
    $<$ 1 ms & $<$ 227 km\\
    \hline
    $<$ 10 ms & $<$ 2270 km\\    
    \hline
    $<$ 20 ms & $<$ 4540 km\\
    \hline
    $<$ 40 ms & $<$ 9080 km\\
    \hline
\end{tabular}
\caption{The maximum distance required to achieve a specific latency.}
\label{tab:minimum_distance}
\end{table}

Table \ref{tab:minimum_distance} shows the projected maximum distance between a server and mobile devices required to achieve a specific latency. As we see, to provide a latency of 1 ms the distance between servers and mobile devices should not exceed 227 km. Hence, a network of Edge servers, distance between which does not exceed 227 km, is required to enable 5G applications, such as Autonomous driving vehicles (see Section \ref{sec:discussion}), demanding 1 ms latency.  
Importantly, this projection does not include the latency introduced by mobile devices and base stations to handle data. Thus, we expect that the maximum distances provided in Table \ref{tab:minimum_distance} may be even smaller.

\begin{table}[t]
\scriptsize
\centering
\begin{tabular}{|c|c|c|c|c|} 
    \hline 
    \textbf{Op1(4G)} & \textbf{Op2(4G)}& \textbf{Op2(5G)} & \textbf{Op3(4G)} &\textbf{Op3(5G)}\\
    \hline
    20.7 ms & 35.1 ms & 27.8 ms & 20.7 ms& 24.3 ms\\
    \hline
\end{tabular}
\caption{The minimum latency measured for different operators.}
\label{tab:minimum_latency}
\end{table}

Finally, Table \ref{tab:minimum_latency} shows the minimum absolute latency which can be provided by each operator. We see that the minimum absolute latency of 20.7 ms is provided by \textit{operator 1} and \textit{operator 3} for 4G LTE. Remarkably, this latency is 17\% lower than the minimum latency achieved on 5G by \textit{operator 3}. Additionally, it outperforms the minimum latency of both 4G LTE and 5G provided by \textit{operator 2} by 70\% and 34\%, respectively.
Similar to our previous results, we attribute this difference to the current limitations within the state-of-the-art 4G/5G ecosystem.

\textbf{Observation6:} \textit{The average latency of the video down-streaming performed by servers located in London is 35 ms and 37 ms for 4G LTE and 5G, respectively.}

\textbf{Observation7:} \textit{The minimum absolute latency which is achieved on a mobile device in London is 20.7 ms and 24.3 ms for 4G LTE and 5G, respectively.}

\textbf{Implication1:} \textit{In addition to the geo-graphical location, the 4G/5G ecosystem can contribute to latency increases of up to 2 times.}

\textbf{Implication2:} \textit{Enabling 1ms latency will require  a network of Edge servers, distance between which does not exceed 227 km.}

\textbf{The impact of routing on latency.}
The down-streaming latency can be significantly impacted by routing and the number of hops. To quantify this effect, we conduct two correlation evaluations. In the first evaluation, we examine the correlation between the number of hops required to reach a server with a specific IP address and the down-streaming latency measured using \textit{ping}. In the second evaluation, we perform a similar correlation using latencies measured with QUIC.

\begin{figure}[H]
\centering
\includegraphics[width=0.8\columnwidth, keepaspectratio]{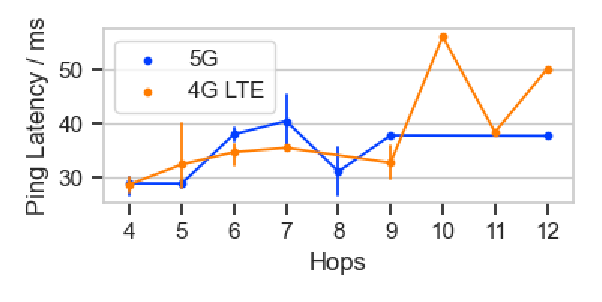}
\caption{The correlation between ping latency and hops.}
\label{ping_hop_lat}
\end{figure}

\begin{figure}[H]
\centering
\includegraphics[width=0.8\columnwidth, keepaspectratio]{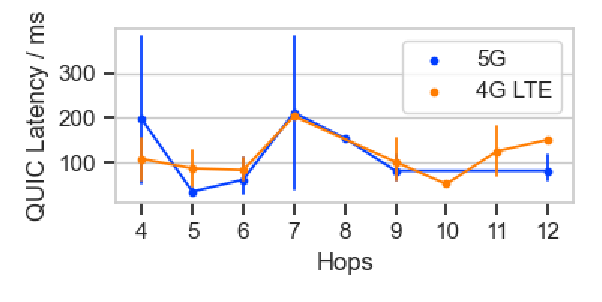}
\caption{The correlation between QUIC latency and hops.}
\label{quic_hop_lat}
\end{figure}

Figure \ref{ping_hop_lat} and Figure \ref{quic_hop_lat} display the correlation between latency and number of hops, averaged over different servers and operators. Interestingly, we observe no strong correlation between the number of hops and latency for both \textit{ping} and QUIC measurements. However, a clear pattern emerges: the average 4G LTE and 5G latencies are at their minimum for 4 hops in the \textit{ping} correlation figure, while the minimum average latencies for 4G LTE and 5G are obtained for 5 hops in the QUIC correlation figure. Moreover, we find that the highest average 5G latency is measured for the servers that can be reached within 4 to 7 hops. These results also suggest that latency is primarily influenced by internal network hardware and configuration in the 4G/5G ecosystem.

\subsection{Bandwidth and energy profiling}
\begin{figure*}
\footnotesize
        \centering
        \begin{tabular}{cccc}
\includegraphics[width=0.45\columnwidth, keepaspectratio]{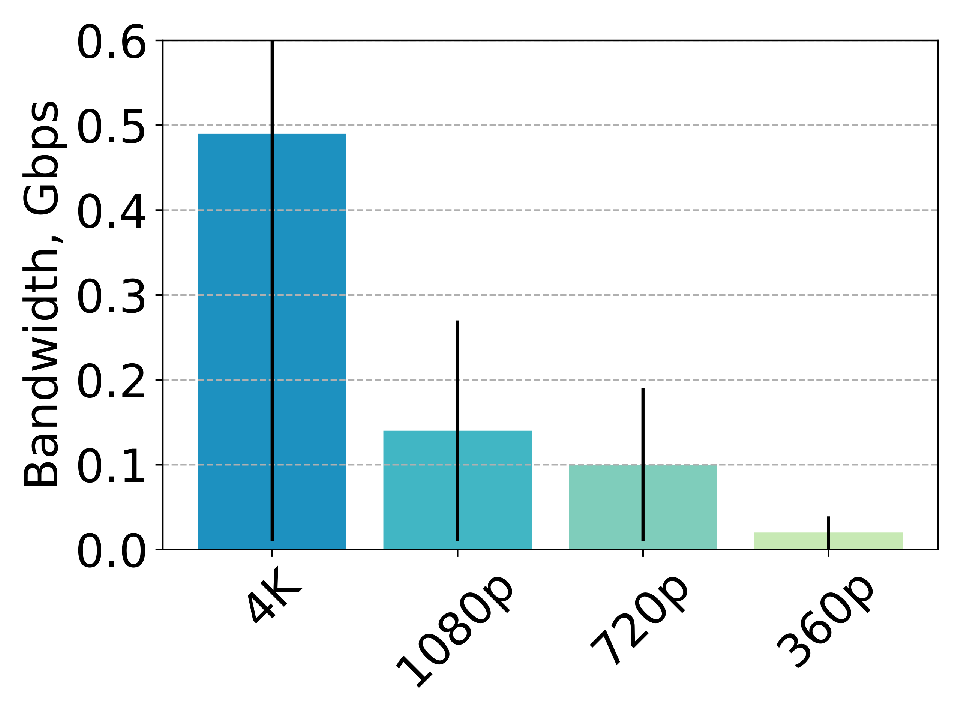} &
\includegraphics[width=0.45\columnwidth, keepaspectratio]{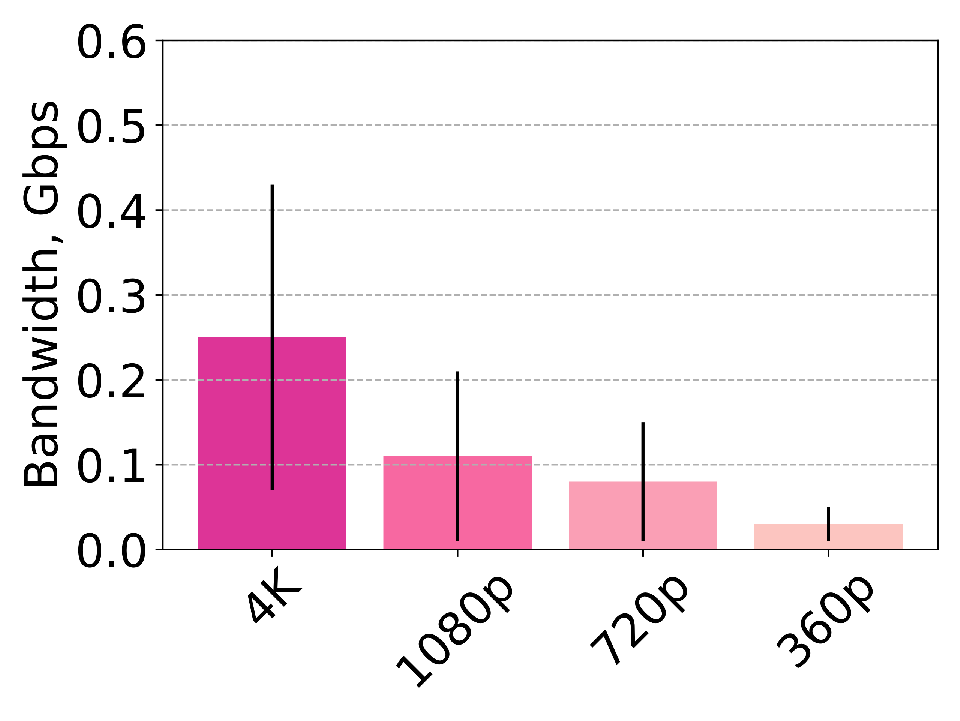} &
\includegraphics[width=0.45\columnwidth, keepaspectratio]{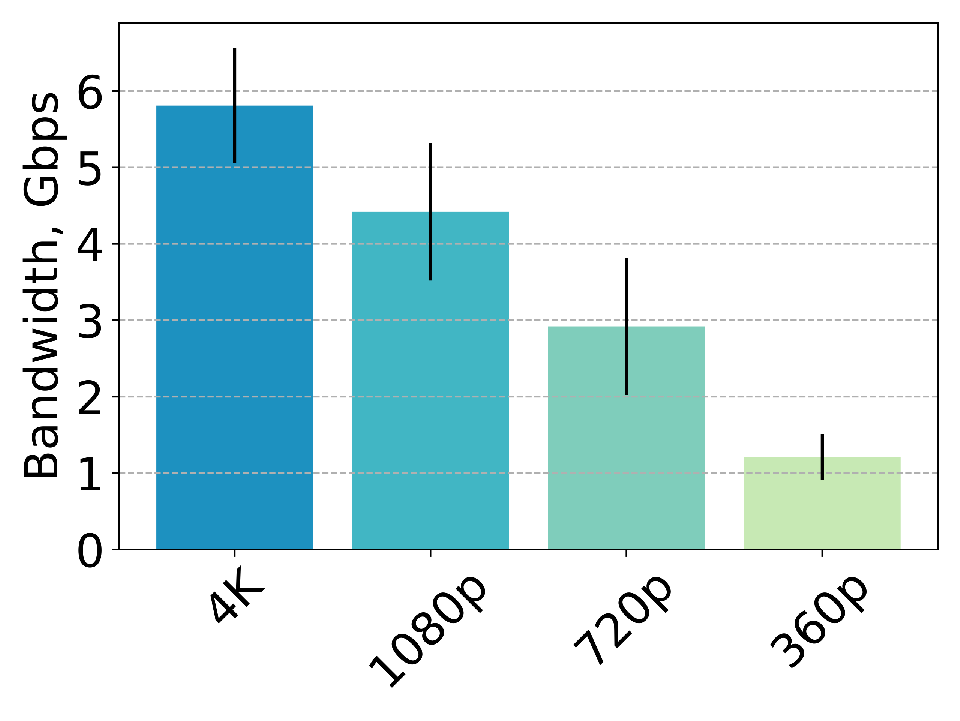} &
\includegraphics[width=0.45\columnwidth, keepaspectratio]{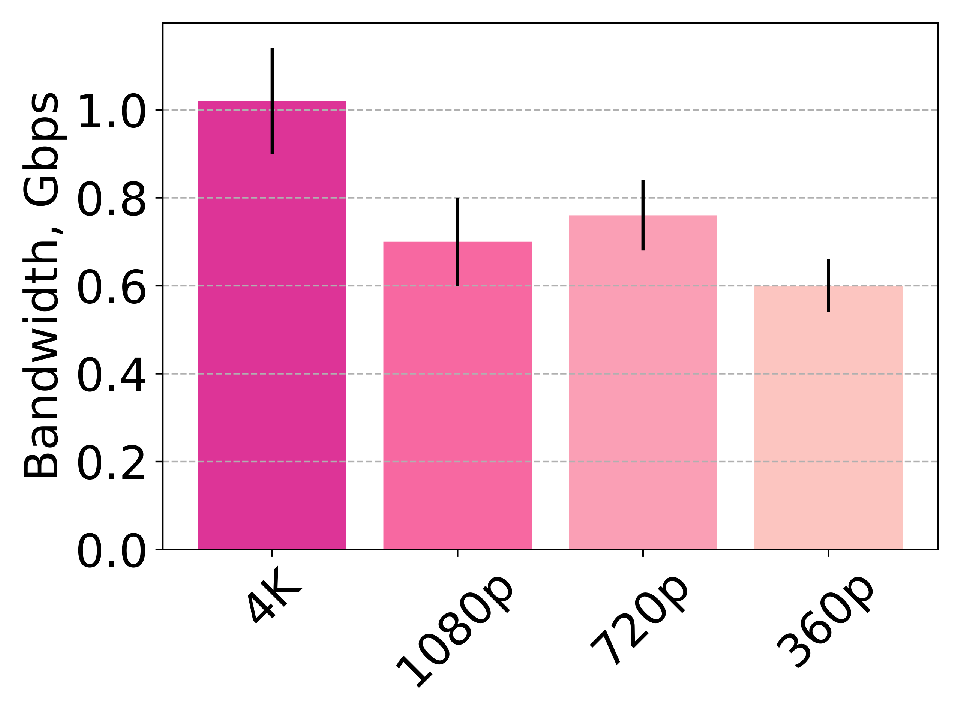}\\
         {A) Bandwidth, 5G} &
         {B) Bandwidth, 4G LTE} &
         {C) Peak Bandwidth, 5G} &
         {D) Peak Bandwidth, 4G LTE} \\
        \end{tabular}
        
        \caption{Average and peak bandwidths for \textit{operator 2}.}
        \label{fig:bandwidth_operator2}
\end{figure*}

\textbf{Bandwidth.} In our next experiments, we measure the bandwidth and device current down-streaming the video with different resolutions. Similar to the previous experiments, we get the bandwidth measurements by parsing the QUIC traces. In our initial set of experiments, we measure the bandwidth using QUIC traces for \textit{operator 2} which provides 5G service with the minimum average down-streaming latency. Figure \ref{fig:table_of_bandwidths} (see Appendix A) shows how the bandwidth measured for 5G and 4G LTE changes when we down-stream the video with 4K, 1080, 720 and 360 resolution. We clearly see that in all the cases the video is down-streamed with some periods and thus we obtain the bandwidth spikes. This is explained by the fact that YouTube streams the video in chunks some of which are buffered in advance, allowing a smoother playback experience. Our second observation indicates that, on average, the amplitude of bandwidth spikes decreases as the video resolution decreases.
We obtain these results for both 4G LTE and 5G. Nonetheless, the amplitude of spikes differs for 5G and 4G LTE. We observe that the bandwidth amplitude is lower for the 4G LTE measurements, however, as expected, the down-streaming periods are longer. To quantify this difference, we averaged the bandwidth for the spikes, i.e. the moments when the video is down-streamed by the YouTube application. Figure \ref{fig:bandwidth_operator2} (A and B) shows the average bandwidth with 95\% confidence interval for the data transfer spikes when we down-stream the video. We see that that the average bandwidth grows with the video resolution and size, and it achieves up to 0.5 Gbps and 0.25 Gbps for 5G and 4G LTE, respectively. Nonetheless, in Figure \ref{fig:table_of_bandwidths}, we see that the peak bandwidth can be much higher for both 5G and 4G LTE. To estimate the peak bandwidth, we parse the traces and sample the highest bandwidth for each period when the smartphone down-streams the videos. Figure \ref{fig:bandwidth_operator2} (C and D) shows the peak bandwidth with 95 \% confidence interval measured for 5G and 4G LTE provided by \textit{operator 2} when down-streaming the videos with different resolutions. Similar to the average bandwidth, we see that the peak bandwidth grows with the video resolution and size. However, the peak bandwidth achieves up to 5.81 Gpbs and 1 Gpbs for 5G and 4G LTE, respectively. Note that these values are higher than the the maximum bandwidth which can be achieved on mid-bands base stations used in our experiments (see Section \ref{sec:exp_setup}). We explain these results later. 

\begin{figure}
\footnotesize
        \centering
        \begin{tabular}{cc}
        \includegraphics[width=0.45\columnwidth, keepaspectratio]{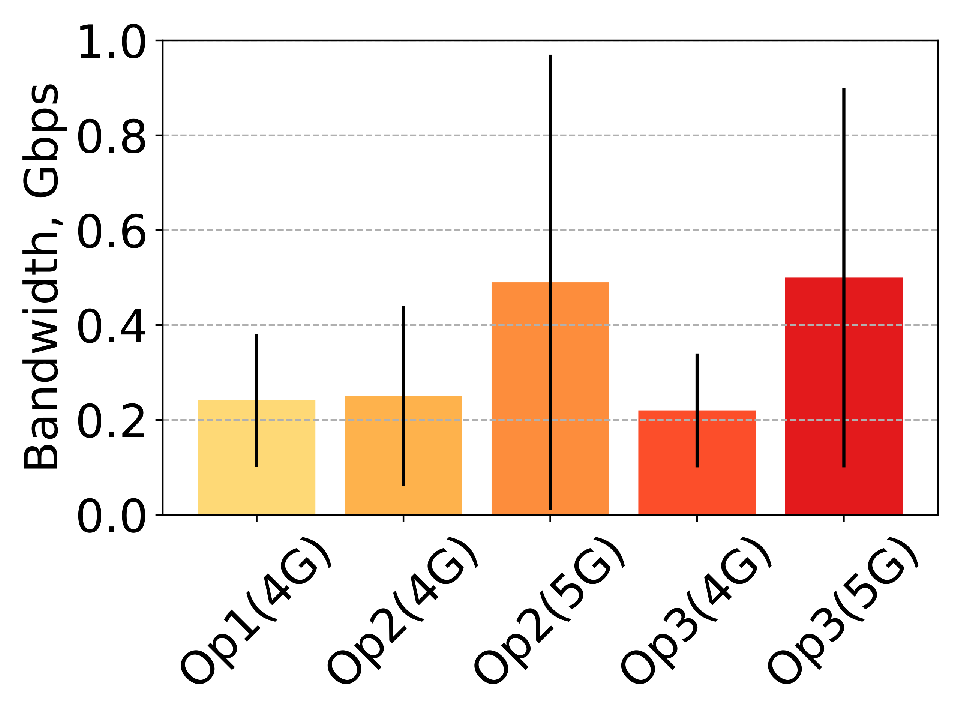} & 
        \includegraphics[width=0.45\columnwidth, keepaspectratio]{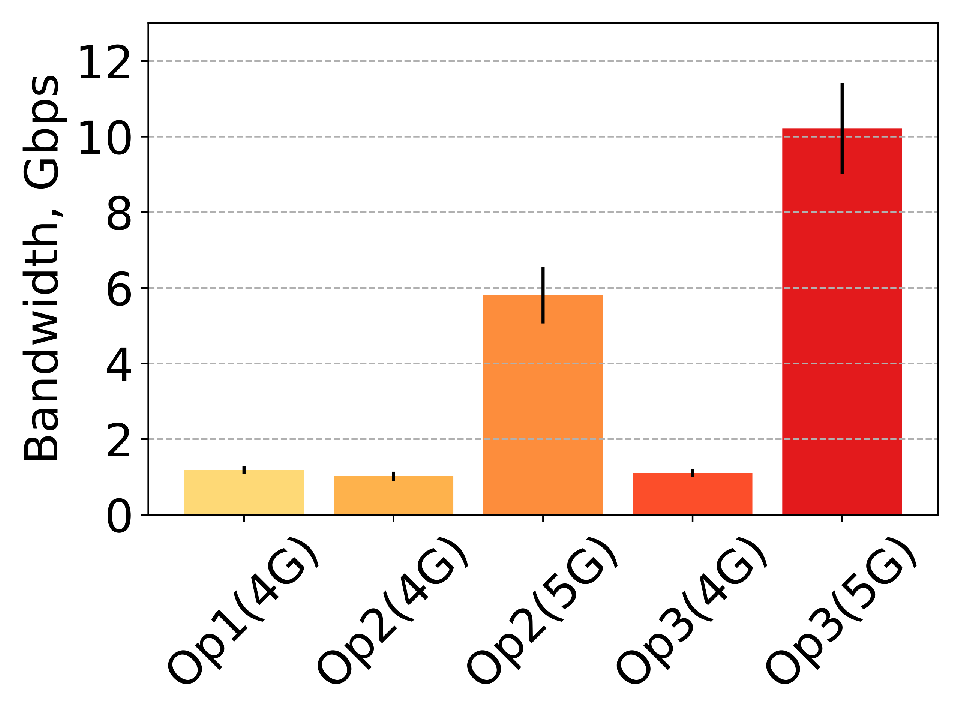}\\    
        {A) Average Bandwidth} &
        {B) Peak Bandwidth} \\                
        \end{tabular}       
        \caption{Average and peak bandwidths across different operators.}
        \label{fig:bandwidth_operators}
\end{figure}

\textbf{Bandwidth across different operators.}
In our experimental study, we also measure the bandwidth across different operators when down-streaming the 4K video. To measure the bandwidth, we employ the same approach as our previous experiments: parsing the QUIC traces and estimating the bandwidth exclusively for down-streaming periods. Figure \ref{fig:bandwidth_operators} (A) shows the average bandwidth with 95\% confidence interval measured for different operators. We observe that the average bandwidth reaches 0.24 Gbps and 0.5 Gbps on average for 4G LTE and 5G, respectively. Thus, we may conclude that the 5G bandwidth is 2$\times$ higher than the 4G LTE bandwidth on average. 

\textbf{Observation8:} \textit{The average down-streming bandwidth for 4G LTE and 5G networks typically range around 0.24 Gbps and 0.5 Gbps, respectively, when considering various network operators.}

We believe that the difference in bandwidth is due to the fact that 5G uses base stations operating at 3.4 Ghz, while 4G uses base stations operating mostly at 2.4 GHz (see Figure  \ref{fig:experiments_map}). These findings can also be attributed to differences in channel bandwidth between 5G and 4G base stations. In 5G base stations used in our study, the channel bandwidth can extend up to 100 MHz, whereas in 4G base stations, the channel bandwidth reaches a maximum of 20 MHz\cite{s22010026,malviya_panigrahi_kartikeyan_2017}. Note that 5G can also benefit from many other optimizations, such as OFDM\cite{7469313}. Interestingly, we notice a consistent average bandwidth among various operators, whereas the average latency exhibits substantial variation across these operators. These findings suggest that latency is more responsive to the networking infrastructure.  

If we compare the peak bandwidths (see Figure \ref{fig:bandwidth_operators}, B), then we observe that the 5G bandwidth is nearly 10 times higher than the 4G LTE bandwidth for \textit{operator 3}, whereas this difference is approximately 6 times for \textit{operator 2}. The highest absolute peak bandwidth, reaching approximately 10 Gbps, is achieved through 5G services offered by \textit{operator 3}. 
Meanwhile, the peak 4G LTE bandwidth is almost the same for all the operators, which is about 1 Gbps. Importantly, 1 Gbps and 10 Gbps are the maximum bandwidths which can be achieved for 4G LTE (Advanced) and 5G networks\cite{bandwidth_maximum}. Nonetheless, the achieved peak bandwidths are higher than the maximum bandwidth, i.e. 0.9 Gbps, which can be achieved on mid-band base stations used in our experiments (see Section \ref{sec:exp_setup}). We explain this by the fact that the 4G/5G modem cannot transfer the received data to CPU instantaneously since CPU reads the data with some delay due to other tasks through a buffer. Thus, CPU may fetch data which has been received previously and the size of fetched data may exceed the maximum size which can be received by the modem per second. These results imply that the mobile CPU may also increase latency.

\begin{figure}
\footnotesize
        \centering
        \begin{tabular}{cc}
        \includegraphics[width=0.45\columnwidth, keepaspectratio]{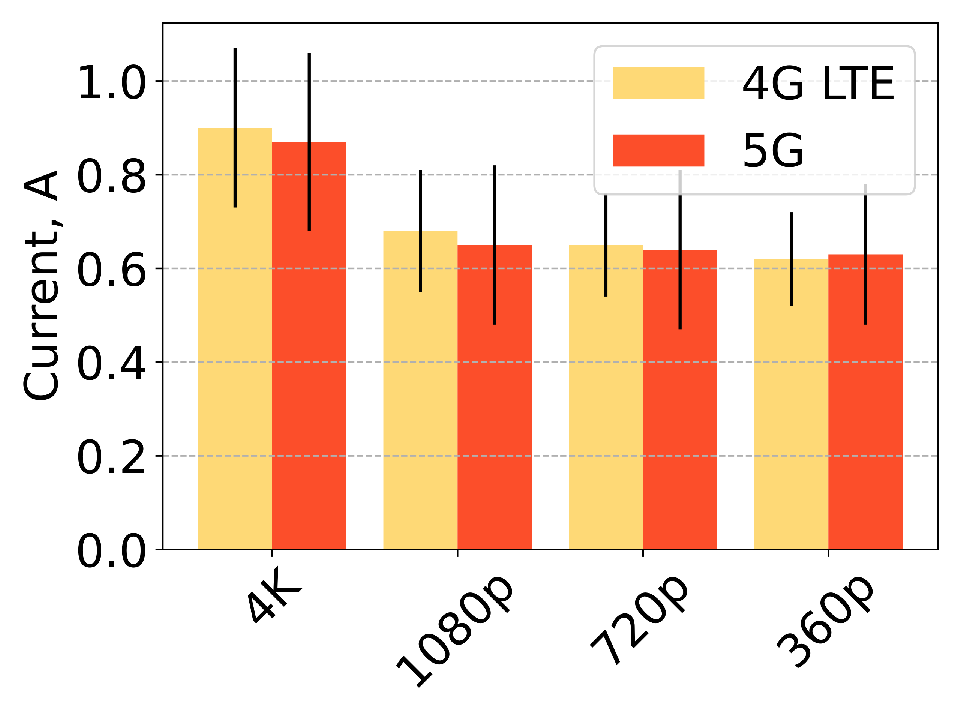} & 
        \includegraphics[width=0.45\columnwidth, keepaspectratio]{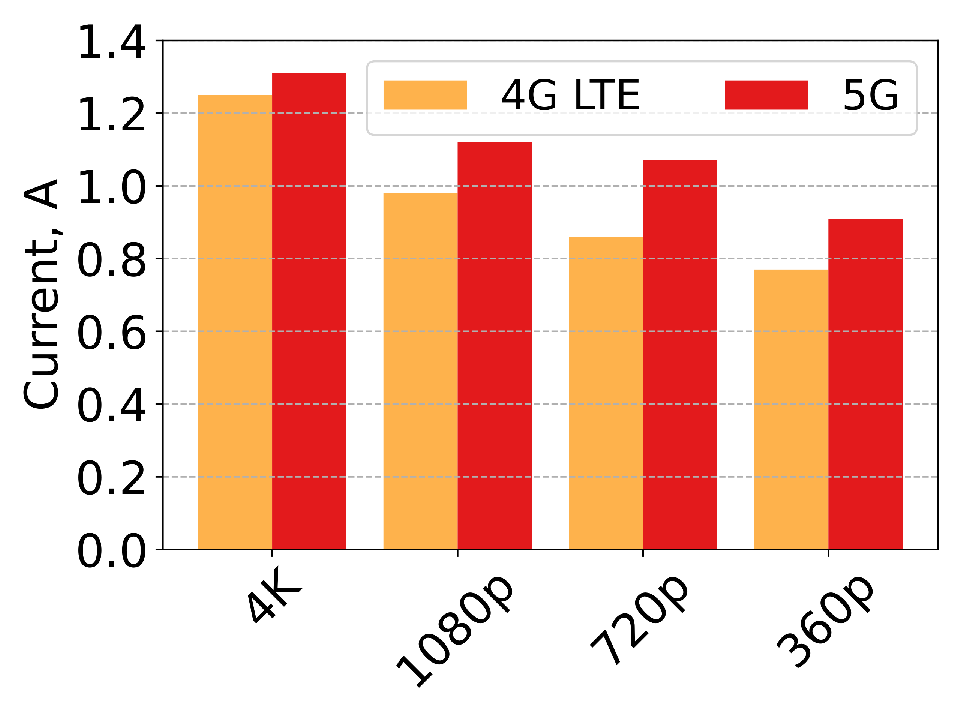}\\    
        {A) Average Current} &
        {B) Peak Current} \\                
        \end{tabular}
        \caption{Device current for \textit{operator 2}.}
        \label{fig:current_operator2}
\end{figure}

\textbf{Power and energy consumption.} To quantify the system power, we measured the mobile device current when down-streaming the videos. Figure \ref{fig:table_of_bandwidths} shows the device current measured for each experiment with bandwidth for \textit{operator 2}. Our first observation is that, similarly to the experiments with bandwidth, we clearly obtain the device current spikes. Moreover, the amplitudes of the device current spikes also decreases when we reduce the resolution of the video. We explain the correlation between the bandwidth and device current spikes by the current which is required to transfer data by the modem. To investigate this, we averaged the current when down-streaming the video. Figure \ref{fig:current_operator2} (A) shows the average current with 95\% confidence interval measured for each down-streamed video. We see that the average current increases by 32 \% for both 5G and 4G LTE  when we down-stream the 4K video  compared to 1080p, 720p and 360p videos. Meanwhile, there is almost no difference between the average current for 1080p, 720p and 360p videos. Note that the size of the video changes with the resolution; the size of 4K, 1080p, 720p and 360p is 578Mb, 178Mb, 142Mb and 28Mb, respectively. Thus, we expect that the average current will grow with the size of transferred data. To investigate this further, we measured the peak current for each experiment with bandwidth, see Figure \ref{fig:current_operator2} (B). Note that we measure the peak current by sampling the highest current values for each period when the video was down-streamed. We see that the peak current grows with the the video resolution, i.e. the size of transferred data, for both 5G and 4G LTE. Moreover, apart from the average current, we clearly see that the peak current is 15\% higher on average for 5G compared to 4G LTE.

\textbf{Observation9:} \textit{ On average, 5G enhances the peak current by approximately 15\% in comparison to 4G LTE.}

This is aligned with the observations made in a previous paper\cite{10.1145/3387514.3405882}. 

\begin{figure}[H]
        \centering
        \includegraphics[width=0.6\columnwidth, keepaspectratio]{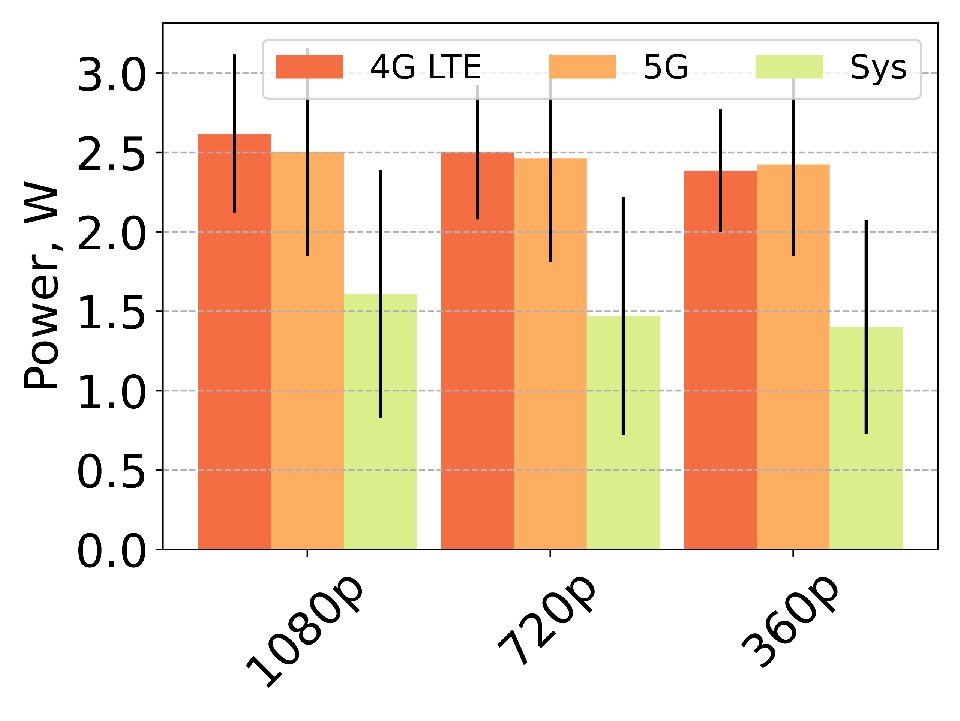}
        \caption{The average energy consumption.}
        \label{fig:energy_measurements}
\end{figure}

To explore the power consumption of a mobile device while streaming videos, we measure current and voltage while playing downloaded video content.
We use the YouTube downloading option to make this experiment. Importantly, the video with 4K resolution cannot be downloaded and played without down-streaming, and thus we make experiments only for 1080p, 720p and 360p videos.
Figure \ref{fig:energy_measurements} shows the average power with 95\% confidence interval estimated using measured current and voltage; in this figure, \textit{Sys} corresponds to the power incurred when playing the videos without down-streaming, i.e. the power incurred by the mobile system without the power induced by down-streaming.

We see that that the down-streaming significantly increases the mobile battery power, by 68\% on average, for both 5G and 4G LTE. These results also imply that the 4G/5G modem consumes 40\% of the total mobile energy on average when down-streaming the data. Thus, the modem receiving the data is one of the biggest contributor to the total mobile energy among screen, CPU, GPU and memory\cite{10.1145/3366423.3380169}.

\textbf{Observation10:} \textit{ The modem down-streaming the data increases the mobile system power by 68\% and consumes 40\% of the total mobile energy.}

Interestingly, there is no significant disparity observed in terms of power and average current between 5G and 4G LTE down-streams. However, 5G radio should consume more power due to more powerful baseband and RF hardware\cite{10.1145/3387514.3405882,7096297}. 
Nonetheless, in our experiments, there is no difference in average power for 5G and 4G LTE. 
We explain this by a higher bandwidth incurred by 5G which implies that the modem spends more time in an idle state without transferring the data when using 5G\cite{10.1145/3387514.3405882}. 
In Figure \ref{fig:table_of_bandwidths} (A and I), it is evident that 5G offers a higher bandwidth in comparison to 4G LTE. However, it is worth noting that 5G transmits data in shorter bursts, while 4G LTE maintains a constant flow of data to the modem, despite having a lower bandwidth. The absence of distinction can also be attributed to mobile operators utilizing 5G Non-Standalone (NSA), which leverages 4G LTE for command data transmission.

\textbf{Observation11:} \textit{On average, the power incurred when down-streaming data is comparable for 5G and 4G LTE.}

\section{Discussion.}
\label{sec:discussion}
\begin{table}[t]
\scriptsize
\centering
\begin{tabular}{|c|c|c|} 
    \hline 
    \textbf{Application} & \textbf{Latency} & \textbf{Bandwidth} \\ 
   \hline
   Mobile cloud gaming \cite{nvidia_cloud_gaming}& \textcolor{ForestGreen}{$<$40ms} & \textcolor{ForestGreen}{$>$45Mbps} \\
   \hline
   Autonomous driving/teleoperation \cite{autonomous_driving}& \textcolor{red}{$~$1ms} & \textcolor{ForestGreen}{$>$100 Mbps} \\
   \hline
   AR/VR(3D rendering) \cite{ar_vr_edge}& \textcolor{red}{$<$20ms} & \textcolor{ForestGreen}{$>$100 Mbps}\\
   \hline
  Unmanned Aerial Vehicles (UAV) \cite{8337920}& \textcolor{ForestGreen}{$~$50ms} & \textcolor{ForestGreen}{50 Mbps} \\ 
   \hline
\end{tabular}
\caption{5G/Edge computing applications.}
\label{tab:edge_applications}
\end{table}

There are many services which can benefit from 5G and Edge computing, however these services have different requirements. For example, Table \ref{tab:location} shows the list of services which are most demanding to latency and bandwidth. 

One of the most promising application for 5G and Edge computing is autonomous driving vehicles\cite{autonomous_driving}. Autonomous driving vehicles should have low latency and high bandwidth connection to enable teleoperation, which is going to be implemented in nearly all commercial autonomous vehicles. It is required for remote supervision, remote assistance and direct operation\cite{autonomous_driving}. However, teleoperation requires a latency of 1 ms, which is $20\times$ lower than the minimum average latency obtained in our experiments. 
Thus, the state-of-the-art 4G/5G ecosystem cannot facilitate a driverless fleet operation in London.

Another potential application is Augmented/Virtual Reality (AR/VR) technology which has not been adopted at scale\cite{ar_vr_edge}. AR/VR promises to enable real-time remote collaboration, guided maintenance and online education to a new level. This technology often uses Edge and Cloud for remote remote rendering. For example, Table \ref{tab:edge_applications} shows the latency and bandwidth requirements, which are 20 ms and 100 Mbps, respectively, for remote 3D rendering used by AR/VR. Thus, our experimental study suggests that the London 4G/5G ecosystem cannot effectively support the implementation of AR/VR technologies, specifically 3D rendering, since the minimum latency obtained in our experiments is 20.71 ms.

Unmanned Aerial Vehicles (UAV) is another potential promising direction which needs 5G and Edge computing. In fact, Amazon started Prime Air delivery in the United States\cite{amazon_air_prime}. To ensure effective operational control of UAVs, 3GPP has established specific requirements for command and control\cite{8434268}. These requirements include data rates of up to 100 Kbps and a latency bound of 50 ms (see table \ref{tab:edge_applications}). Meanwhile, use cases involving flying cameras and remote surveillance rely on UAVs to transmit real-time telemetry data, pictures, or videos. The primary connectivity requirement for such data communication is the data rate, which can reach up to 50 Mbps\cite{8434268}. All these requirements can be met by the 4G LTE and 5G networks tested in our experiments.

\begin{figure}[H]
        \centering
        \includegraphics[width=0.8\columnwidth, keepaspectratio]{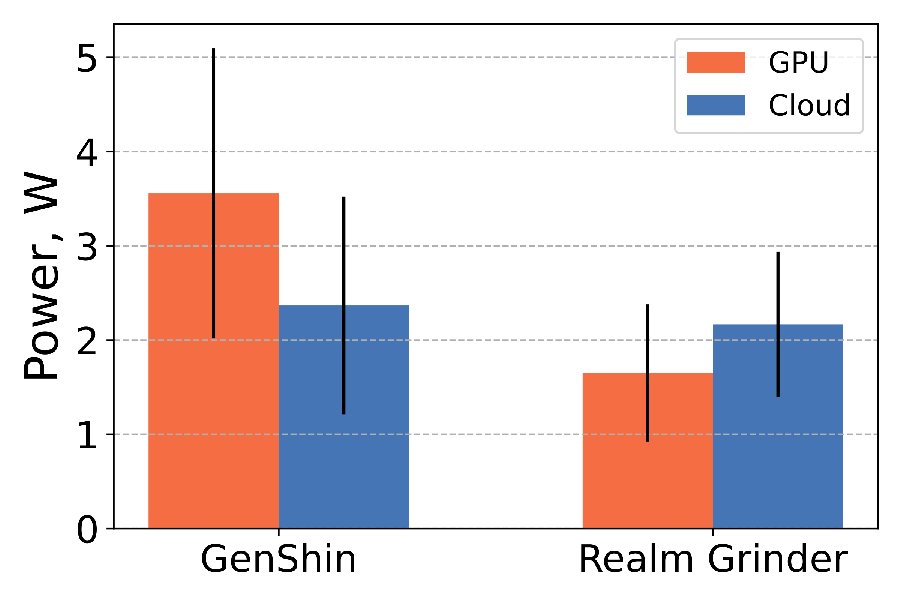}
        \caption{The average power measured for mobile games running rendering on Cloud and local GPU.}
        \label{fig:games_power}
\end{figure}

Mobile cloud gaming is one of the most promising applications of 5G networks for mobile devices \cite{nvidia_cloud_gaming,xbox_cloud_gaming,10.5555/3485849.3485850}. The main idea behind this approach is to offload the GPU graphical pipeline in Cloud for remote rendering and down-stream all the frames. To enable high quality gaming, the network latency and bandwidth should satisfy specific criteria. For example, Nvidia GeForce Now demands a latency of less than 40 ms and bandwidth higher than 45 Mbps for the best experience, i.e. 3840x2160 pixel resolution at 120 FPS\cite{nvidia_cloud_gaming}. Although, the 4G/5G ecosystem tested in our experiments meets this criteria, down-streaming significantly increase the mobile system power, as we show earlier. To investigate this further, we run 2 different games using the Cloud gaming service provided by NVidia Now. To be more specific, we use popular \textit{GenShin Impact} and \textit{Realm Grinder} games. Figure \ref{fig:games_power} shows the average power with 95\% confidence interval incurred when running the games on NVidia Now Cloud and mobile GPU. Note that NVidia Now uses Edge servers placed in London where the latency varies from 37 ms to 145 ms. We see that the GPU version of \textit{GenShin Impact} uses 50\% more power than the Cloud version. However, for \textit{Realm Grinder} we observe opposite results since the Cloud version consumes 31\% more power compared to the GPU version. Such a difference is explained by the fact that  \textit{GenShin Impact} is a social simulation game which uses the maximum capacity of GPU, while \textit{Realm Grinder}  is an RPG game which does not use GPU heavily. Note that \textit{GenShin Impact} is one of the most demanding mobile games which uses the OpenGLES interface, while many games these days use Vulcan which requires significantly less power\cite{9605447}. Overall, based on these results, we may conclude that offloading GPU workloads into Cloud/Edge may be inefficient.
Moreover, the use of Cloud and Edge facilities may be inefficient for other applications, such as AR/VR and Unmanned Aerial Vehicles. Thus, apart from bandwidth and latency, energy efficiency of offloading computing in Cloud/Edge should be considered as an important challenge for future 5G applications, especially for those applications where energy consumption is critical.

\textbf{Implication3:} \textit{Offloading computation to Cloud/Edge presents a significant energy consumption challenge, as local computing units can prove more energy-efficient than transmitting computation results.}

Expanding the utilization of Cloud/Edge technology for down-streaming on mobile devices may also inadvertently contribute to a rise in global carbon emissions. Subsequently, we present a projection of the worldwide energy consumption and emissions resulting from smartphones down-streaming YouTube videos.

\textbf{Global energy consumption and carbon emissions due to down-streaming.}
\begin{figure}[H]
        \centering
        \includegraphics[width=0.8\columnwidth, keepaspectratio]{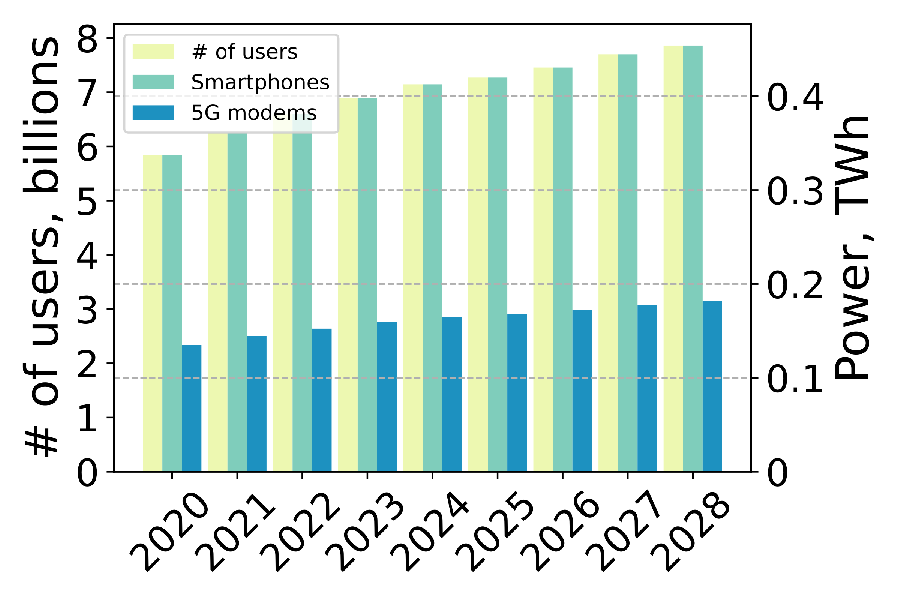}
        \caption{The projected worldwide energy consumption.}
        \label{fig:global_energy}
\end{figure}
There are 6592 million smartphone users today which number grows every year. It is projected that this number will achieve 7861 million users by 2028\cite{number_smartphone_users} (see Figure \ref{fig:global_energy}, right y-axis). It also has been estimated that each smartphone user spends 23.1 hours per month watching YouTube on average\cite{wathing_youtube_number}. Assuming that each user watches YouTube videos at 720p resolution, we can estimate the global energy consumption due to watching YouTube on smartphones using the average Pixel power consumption measured in our experiments for the 720p video, the average time spent watching YouTube per month and the number of smartphone users. Figure \ref{fig:global_energy} (left y-axis) shows the total global energy spent by smartphones (\textit{Smartphones}) and 5G modems (\textit{5G modems}) when down-streaming YouTube videos at 720p resolution. Note the total power estimated will be almost the same for 360p resolution videos according to the power measurement taken for the mobile device used in our experiments (see Figure \ref{fig:bandwidth_operators}). In these projections, we assume that the average power consumption of smartphones, including modem power, is equivalent to that of the Google Pixel 4a used in our experiments. 

We see that the smartphone users spent almost 0.4 TWh per month on average in 2022 on watching YouTube videos, while this number will exceed 0.45 TWh by 2028. Note that almost 40\% of this energy, which is about 0.15 TWh in 2022, is spent by mobile modems to receive the data from Cloud/Energy. As per the United States Environmental Protection Agency (EPA), the utilization of 0.4 TWh and 0.15 TWh corresponds to emissions equivalent to 73038 tons and 64889 tons of carbon dioxide ($CO_{2}$), respectively\cite{epa}. Moreover, 0.43 TWh is the amount of energy produced per month by R.E. Ginna reactor\cite{reactor_energy}. Thus, to facilitate only global mobile YouTube down-streaming and watching on smartphones, the infrastructure equivalent to an entire nuclear power plant is necessary. Moreover, we projected only the energy consumption for smartphones, however the total amount of energy spent on streaming videos is much higher due to base stations and Cloud/Edge data centers, as previous studies show\cite{netflix_energy}. Hence, the worldwide energy consumption resulting from mobile device usage for streaming YouTube videos, as well as accessing other online services, and its impact on global $CO_{2}$ emissions is significantly greater. 

Thus, when developing 4G/5G applications that utilize Cloud and Edge technologies, giving paramount importance to energy efficiency is crucial not only for managing energy consumption but also for mitigating global carbon emissions.

\textbf{Summary.}
Overall, our study reveals that the 5G ecosystem is currently in its early implementation stage. While 5G offers a bandwidth that is 2 times higher than 4G LTE, it is noteworthy that certain operators still demonstrate lower latency with their 4G LTE networks compared to 5G. We attribute this disparity to the inadequacy of the 5G ecosystem in London, including base stations, Cloud/Edge servers, and interconnection networks, which are not yet prepared to deliver low-latency data transfers. Consequently, services requiring an ultra-low latency of 1 ms, such as Autonomous driving/teleoperation, are unable to leverage the benefits of 5G in London these days. 

To achieve an ultra-low latency and average bandwidth close to 10 Gbps, it is essential to deploy high-band base stations and cells supporting mmWaves throughout London. This endeavor necessitates substantial investment because mmWaves base stations have significantly shorter effective ranges compared to mid-band stations\cite{8647379}. Furthermore, as demonstrated in this study, achieving a latency of 1 ms will necessitate a network of Edge servers positioned no more than 227 km apart. 

Another pressing concern revolves around the energy efficiency of mobile devices and 4G/5G modems, rendering certain services impractical, such as Mobile Cloud gaming or remote AR 3D rendering, which heavily rely on offloading computing to Cloud/Edge. Therefore, aside from enhancing the 5G ecosystem and implementing new wireless 6G standards, it is imperative to prioritize optimizing the energy efficiency of mobile devices, especially 4G/5G modems. This task holds increasing significance as the number of devices utilizing mobile 5G networks continues to grow, alongside the escalating mobile Internet traffic, projected to reach 329 Exabytes per month by 2028\cite{global_intetnet}.
\section{Limitation of our study.}
\label{sec:limitation}
One of the key requirements for emerging 5G applications, such as Autonomous Driving Vehicles and telesurgery\cite{Xu2014DeterminationOT,Ebihara2022TeleassessmentOB} is reliability. However, unfortunately, in our study we were not able to measure the packet loss rate since QUIC encrypts this information, which we were not able to decrypt. Nonetheless, a prior study revealed that when utilizing UDP for data transfer, the packet loss rates can reach up to 4\% for 5G and 0.9\% for 4G LTE, particularly under conditions of high bandwidth \cite{10.1145/3387514.3405882}. Note that 5G URLLC ( Ultra Reliable and Low Latency Communications) used for Autonomous Driving Vehicles requires 0.0001\% packet error rate. Thus, apart from latency, 5G reliability also prevents a driverless fleet operation in London.  

In our experiments, we use a stationary position for the experiments. It is important to acknowledge that prior research has established a correlation between the separation distance of a smartphone from base stations and degradation of latency and bandwidth\cite{10.1145/3387514.3405882,10.1145/3452296.3472923}. The impact of structures and atmospheric conditions further influences these metrics \cite{9987496,electronics9111867,https://doi.org/10.1002/ett.3311,8641429}. Nevertheless, it is crucial to emphasize that our study's primary objective revolved around understanding the minimum latency and maximum bandwidth that can be achieved when the distance between the smartphone, base stations and Edge servers is minimized. 

In our study, we exclusively focus on down-streaming scenarios, despite the fact that mobile network performance varies between down-streaming and up-streaming activities. Past research has shown that the up-link bandwidth is notably lower than the down-link bandwidth, while energy consumption remains relatively consistent \cite{10.1145/3452296.3472923}. Meanwhile, our study aims to comprehend the attainable maximum performance of real-world applications on commercial mobile 5G networks.
\section{Conclusion}
\label{sec:concl}
This paper presents the findings of our experimental study conducted in London, which aims to understand what latency and bandwidth for 4G LTE and 5G can be achieved on a real commercial device using Edge servers in practice. Our study reveals that despite the 5G networks providing an average bandwidth up to 2 $\times$ higher than 4G LTE, the 4G LTE networks in London exhibit lower latency compared to their 5G counterparts. 
These results are explained by the inadequacy of the 5G ecosystem in London, including base stations, Cloud/Edge servers, and wired/fiber networks, which implementation is at the early stage. Our study demonstrates that the 4G/5G ecosystem can contribute to latency
increases of up to 2 times.
Thus, the 4G/5G ecosystem is a major bottleneck which prevents exploiting full capabilities of 4G LTE and 5G networks. Additionally, our study uncovers a critical issue with the energy consumption of mobile devices, particularly 4G/5G modems, which contribute to 40\% of the total mobile energy consumption. We demonstrate that the combination of high latency and significant energy consumption poses obstacles to leveraging 5G for crucial applications like Cloud gaming and Autonomous driving/teleoperation.

We firmly believe that to fully unlock the potential of 5G, it is essential to prioritize efforts aimed at improving the 5G ecosystem and enhancing the energy efficiency of mobile devices.

\clearpage
\newpage
\onecolumn
\appendix
\section{Appendix}
\begin{figure}[H]
\footnotesize
        \centering
        \begin{tabular}{cc}
            \includegraphics[width=0.5\columnwidth, keepaspectratio]{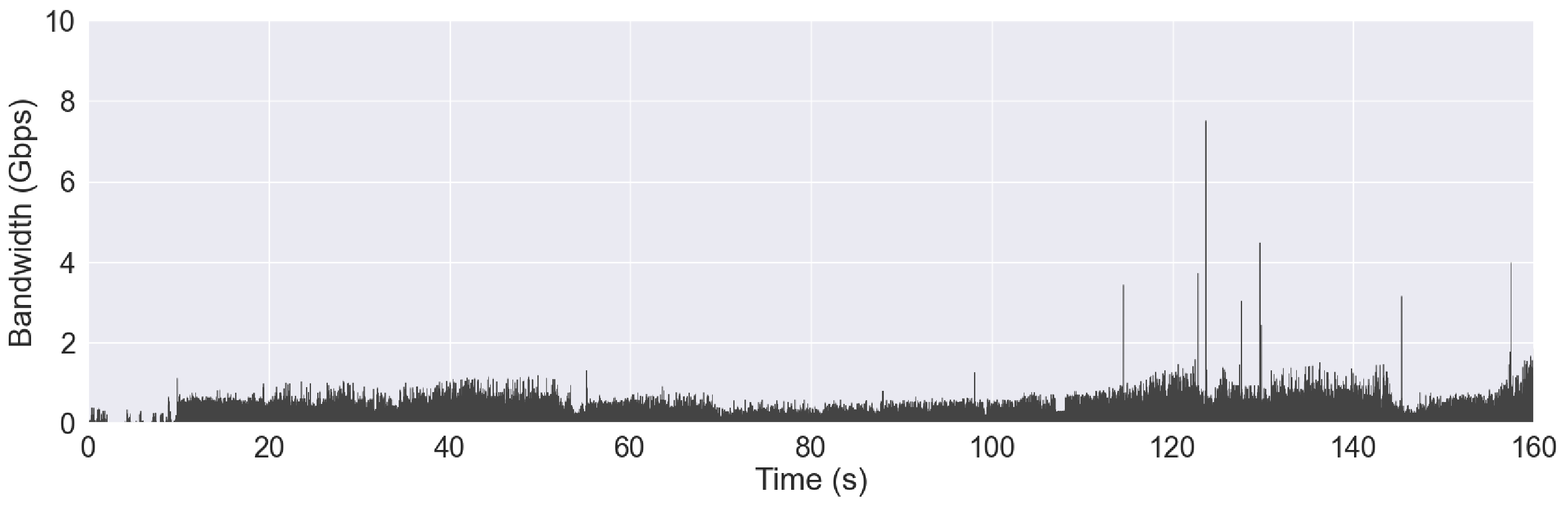} & \includegraphics[width=0.5\columnwidth, keepaspectratio]{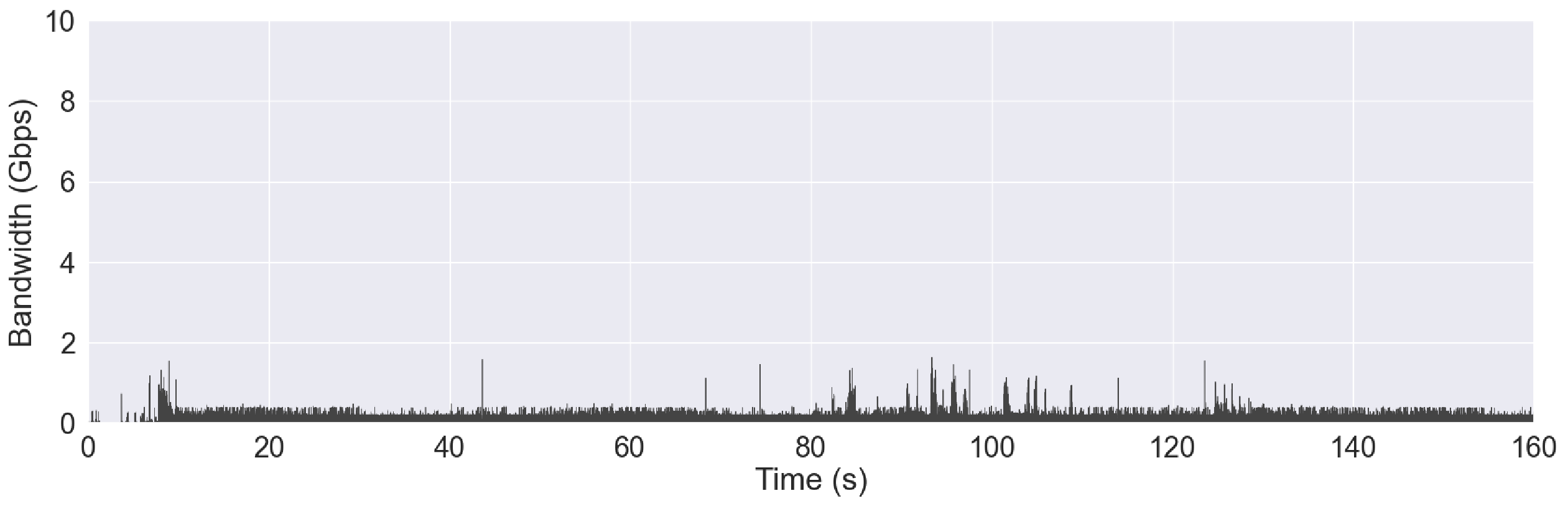}\\
            Bandwidth (Cloud Genshin) & Bandwidth (Cloud Realms Grinder)\\
            \includegraphics[width=0.5\columnwidth, keepaspectratio]{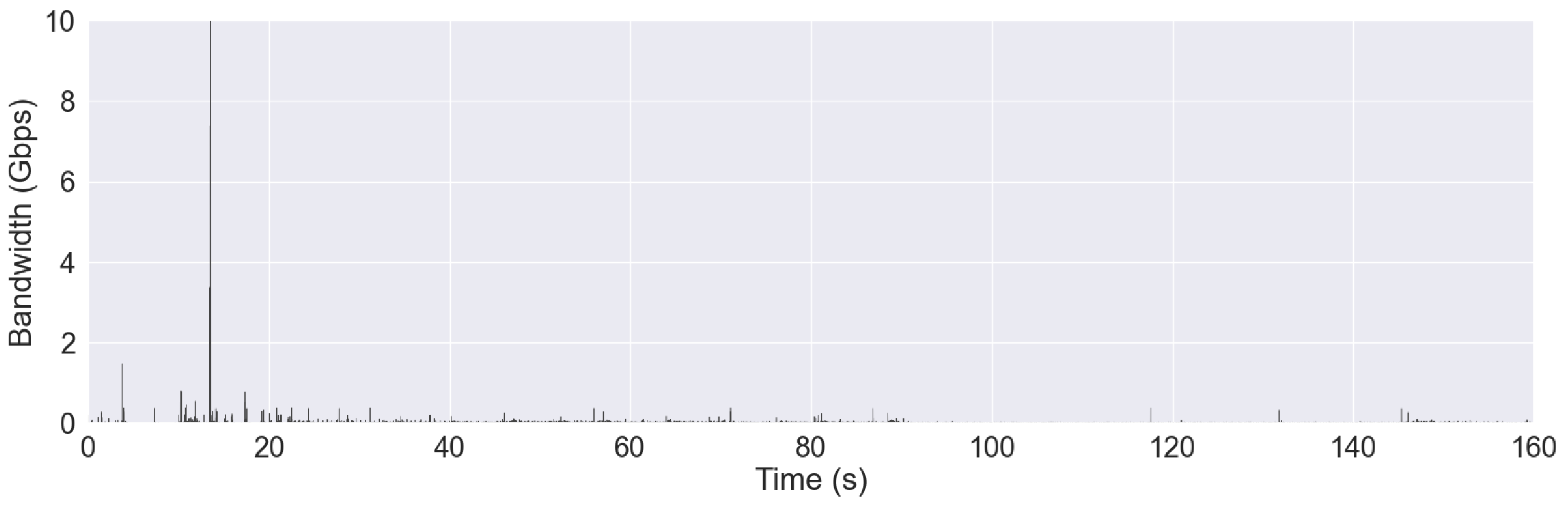} & \includegraphics[width=0.5\columnwidth, keepaspectratio]{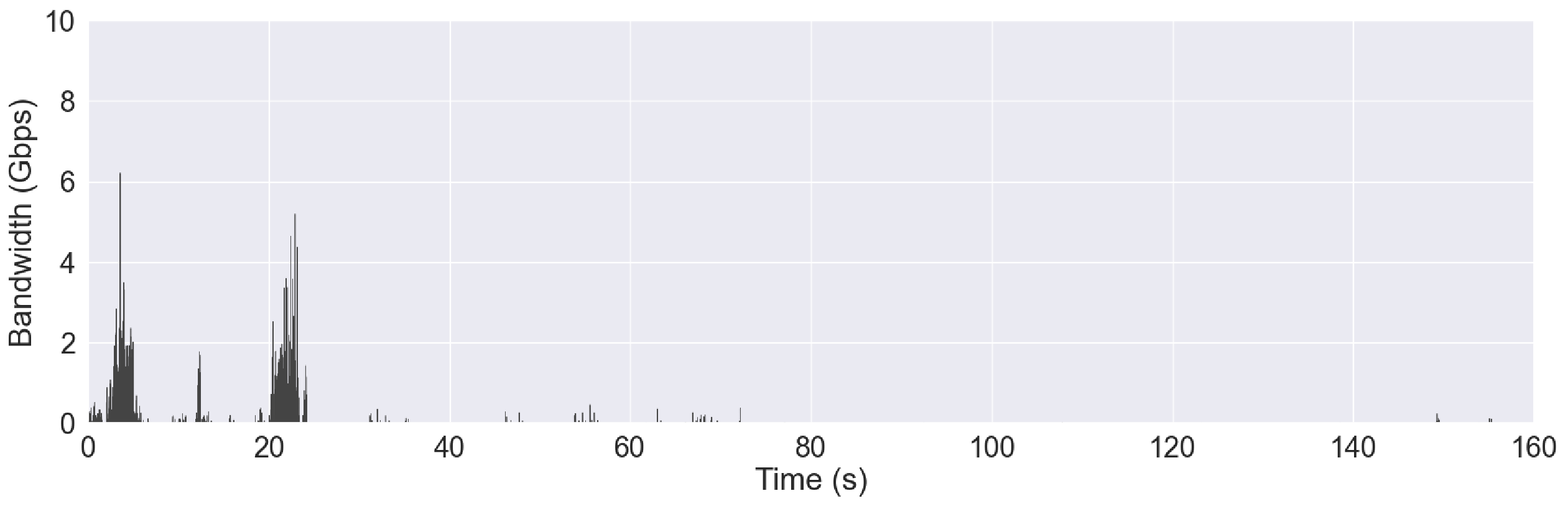} \\
            Bandwidth (Genshin) & Bandwidth (Realm Grinder)\\
            \includegraphics[width=0.5\columnwidth, keepaspectratio]{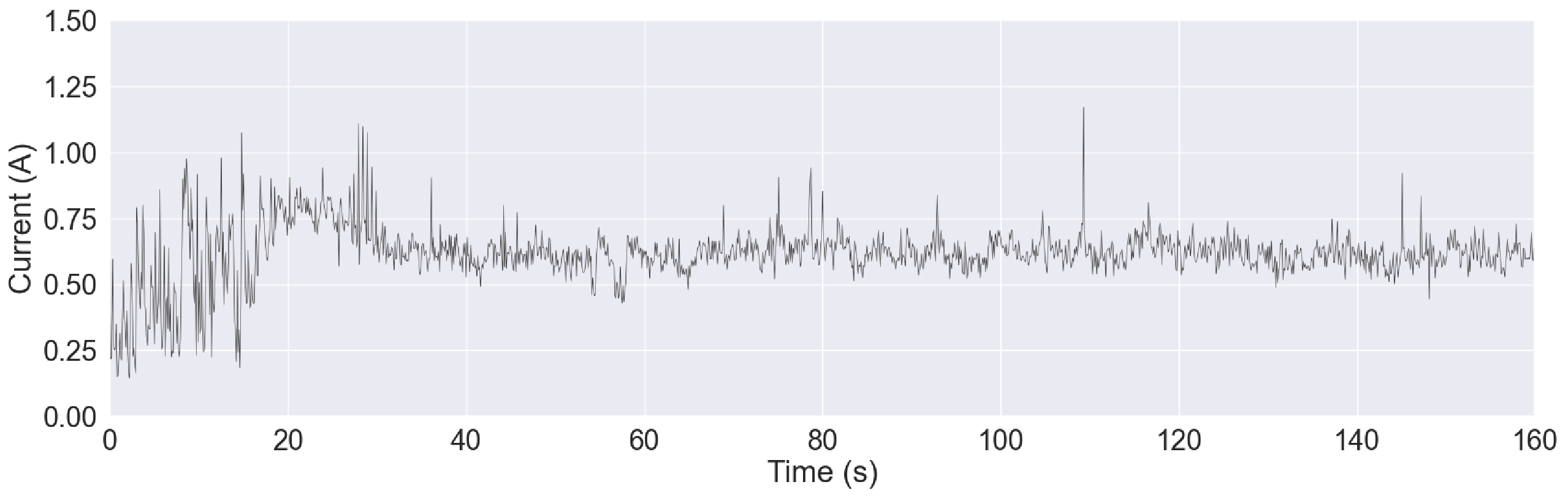} & \includegraphics[width=0.5\columnwidth, keepaspectratio]{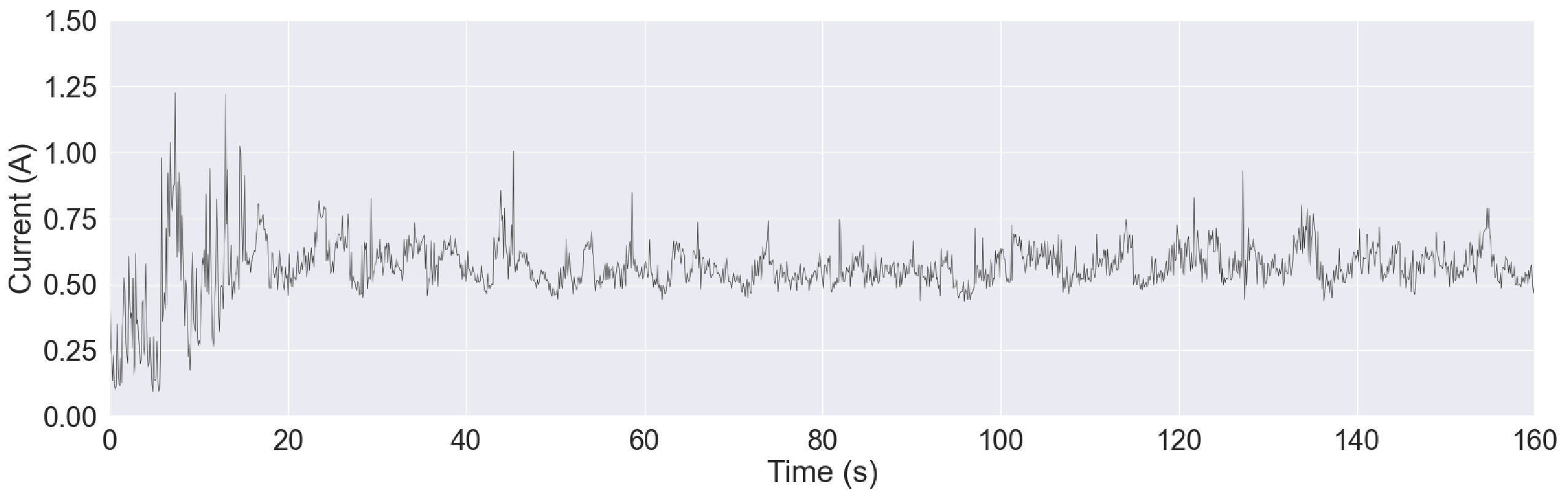}\\
            Current (Cloud Genshin) & Current (Cloud Realm Grinder)\\
            \includegraphics[width=0.5\columnwidth, keepaspectratio]{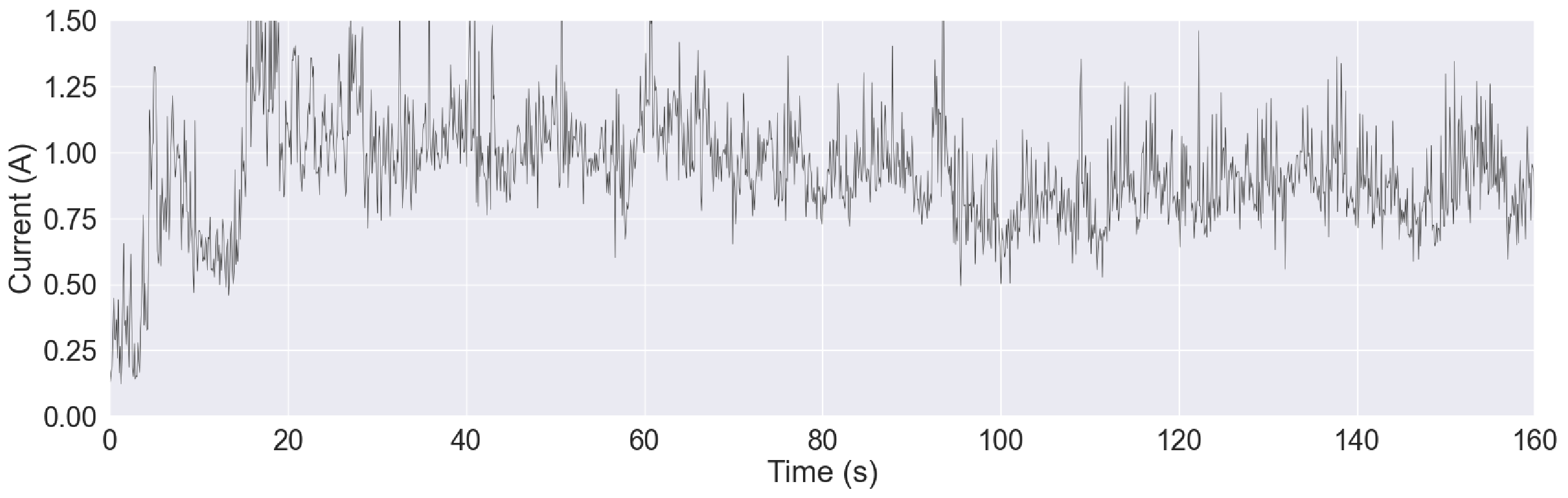} & \includegraphics[width=0.5\columnwidth, keepaspectratio]{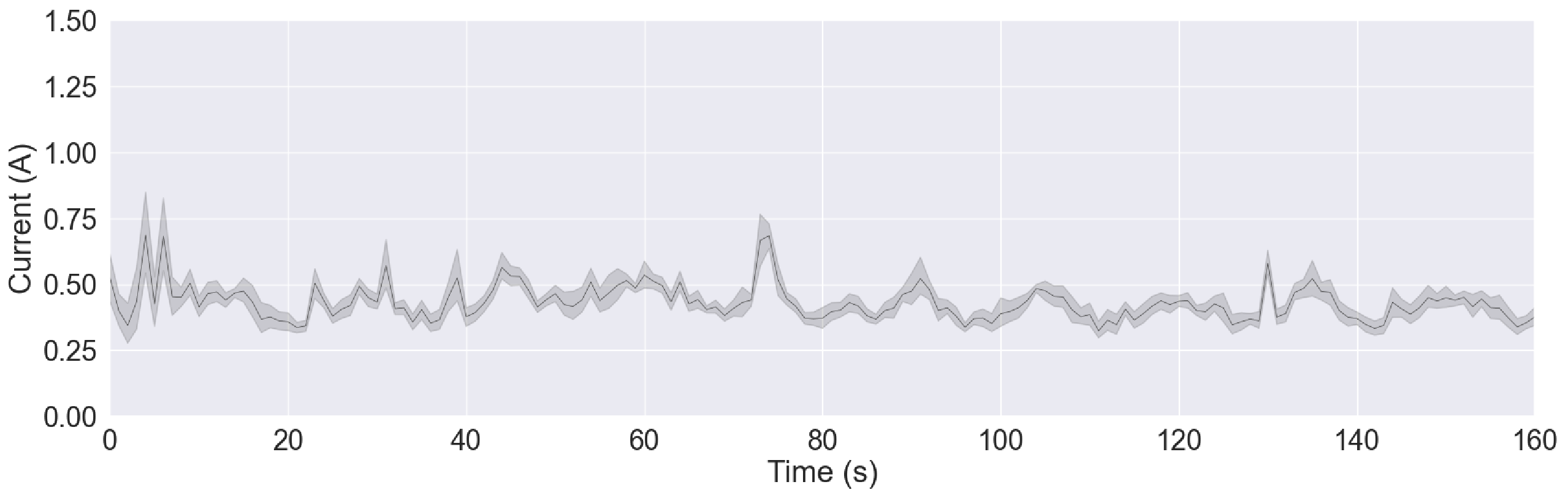} \\
            Current (Genshin) &  Current (Realm Grinder)\\
        \end{tabular}
        \caption{Bandwidth and current with and without cloud gaming}

        \label{fig:genshin_profiling}
\end{figure}

\begin{figure*}[ht]
\footnotesize
        \centering
        \begin{tabular}{cc}
            \includegraphics[width=0.5\columnwidth, keepaspectratio]{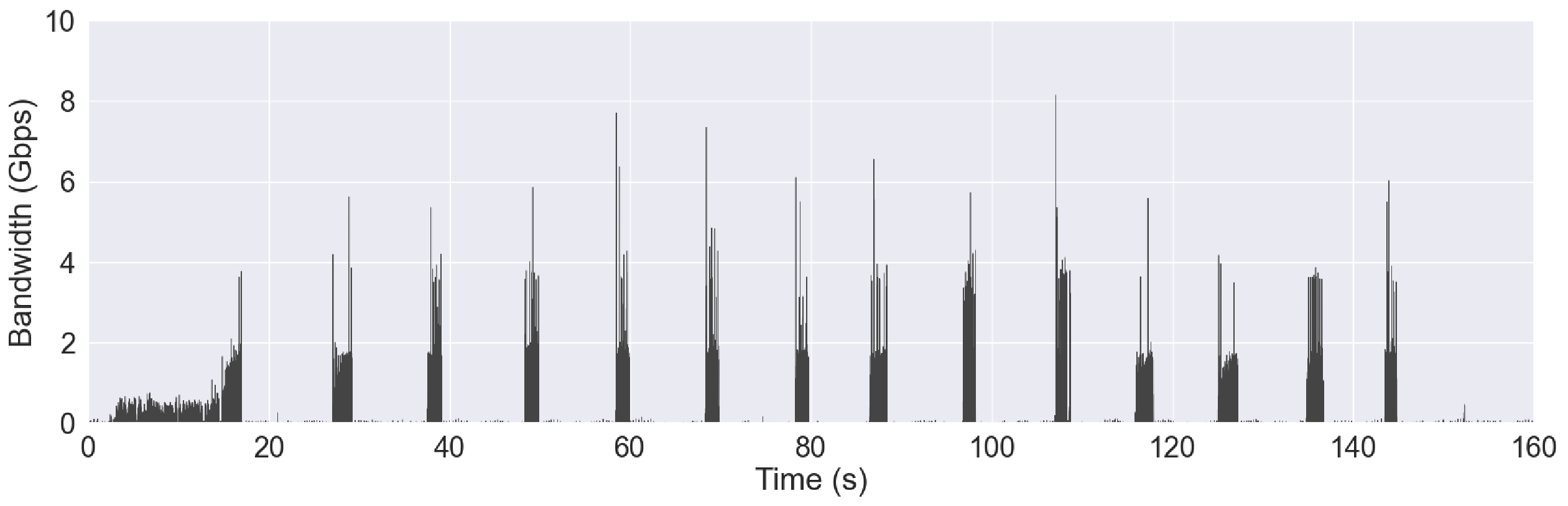} & 
            \includegraphics[width=0.5\columnwidth, keepaspectratio]{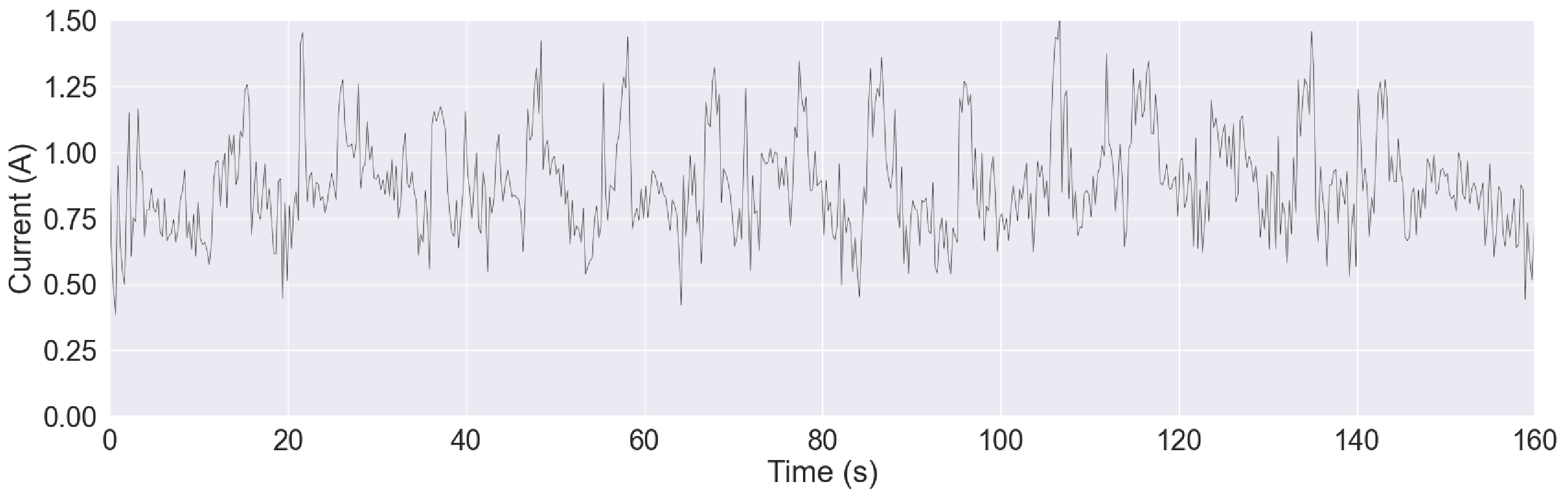} \\    
            {A. Bandwidth, 4K resolution, 5G } &
            {B. Current, 4K resolution, 5G}\\
            \includegraphics[width=0.5\columnwidth, keepaspectratio]{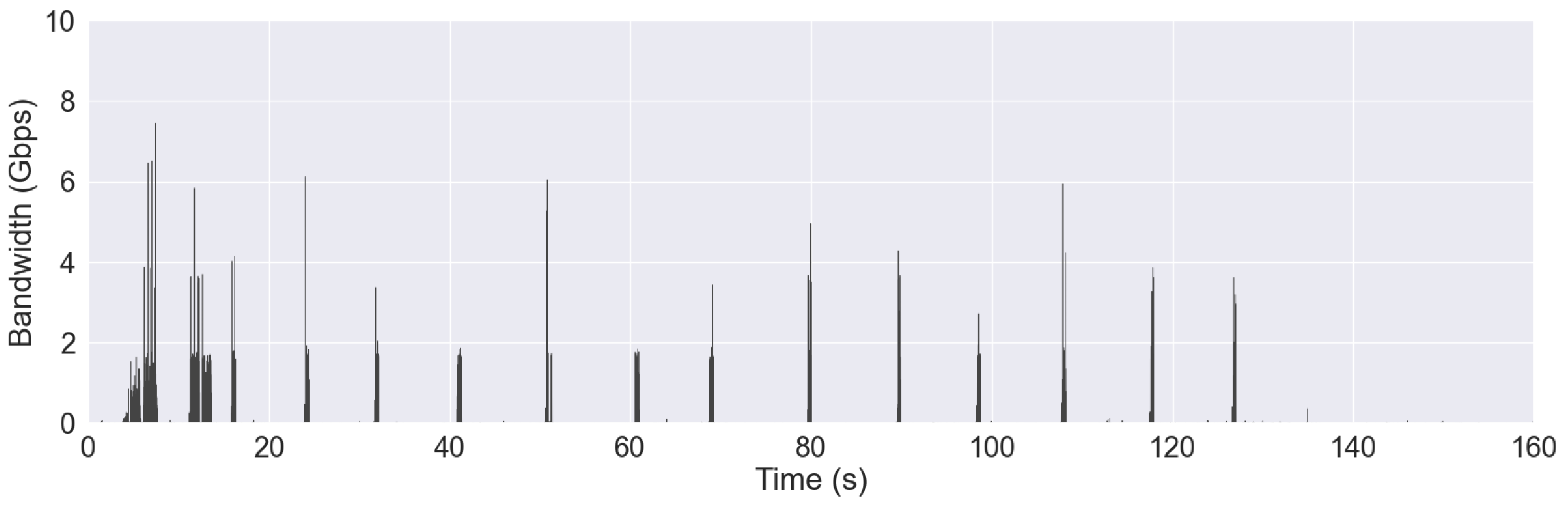} &
            \includegraphics[width=0.5\columnwidth, keepaspectratio]{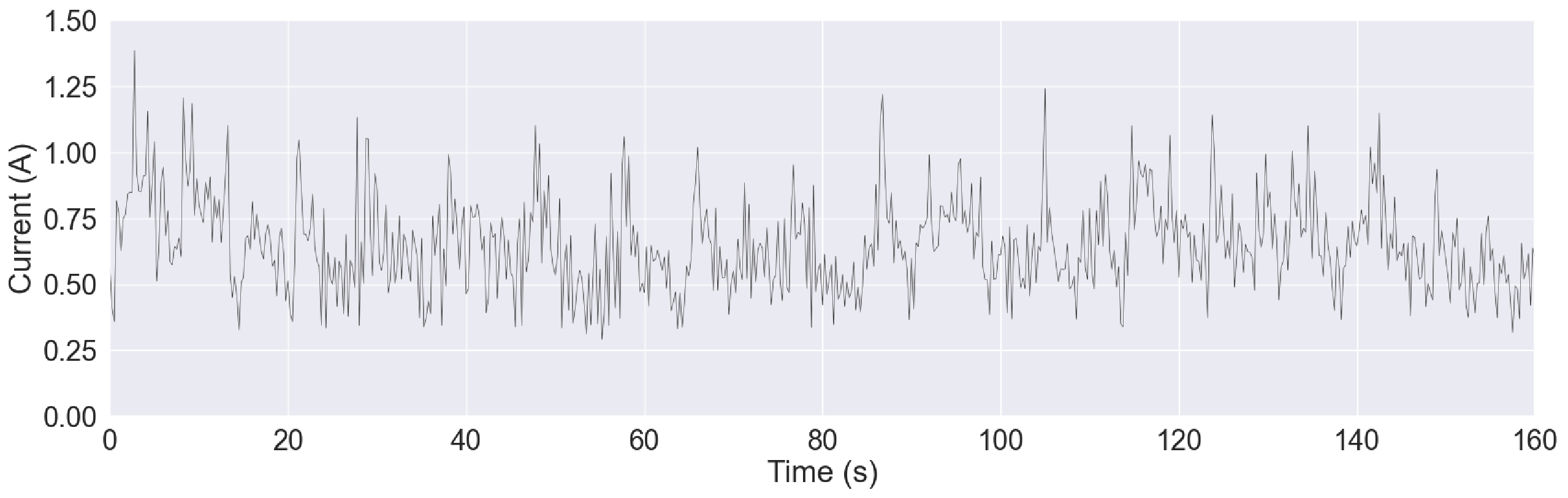}\\
            {C. Bandwidth, 1080p resolution, 5G} &
            {D. Current, 1080p resolution, 5G}\\
            \includegraphics[width=0.5\columnwidth, keepaspectratio]{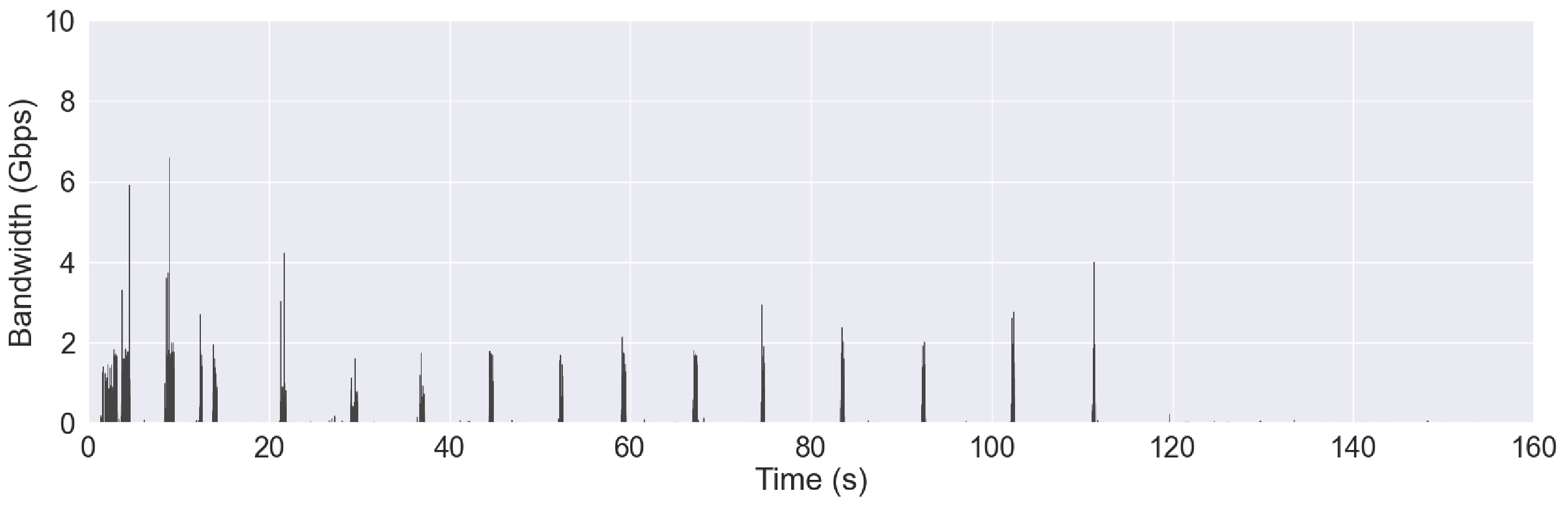} &
            \includegraphics[width=0.5\columnwidth, keepaspectratio]{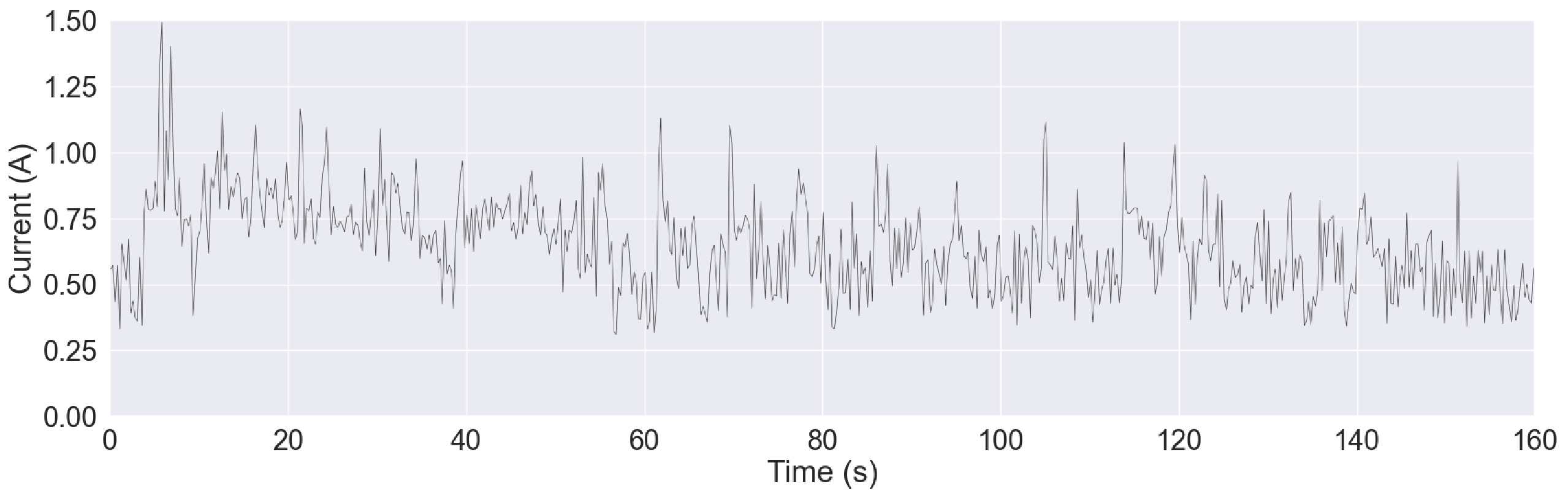} \\
            {E. Bandwidth, 720p resolution, 5G} &
            {F. Current, 720p resolution, 5G}\\
            \includegraphics[width=0.5\columnwidth, keepaspectratio]{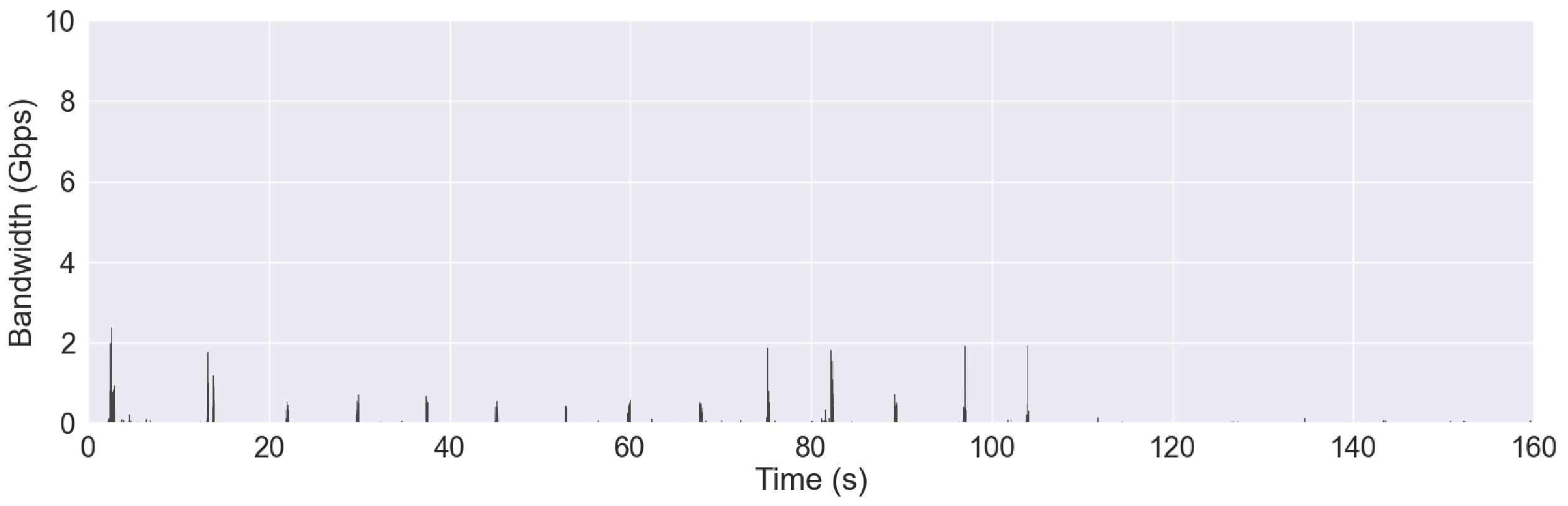} &
            \includegraphics[width=0.5\columnwidth, keepaspectratio]{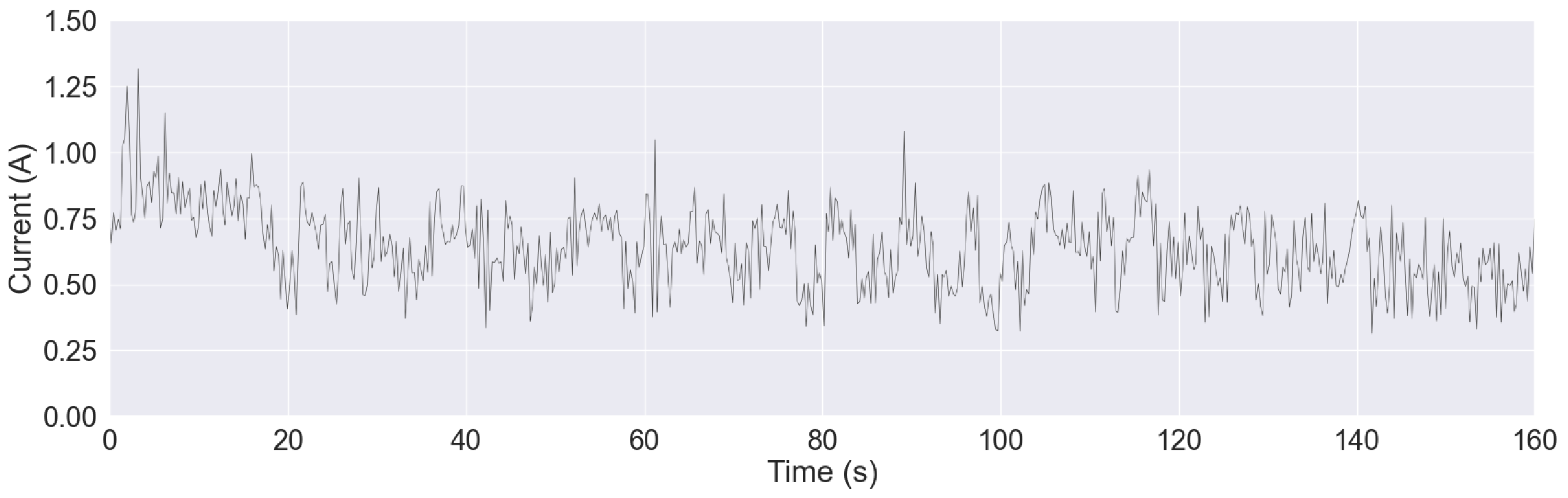} \\
            {G. Bandwidth, 360p resolution, 5G} &
            {H. Current, 360p resolution, 5G}\\

            \includegraphics[width=0.5\columnwidth, keepaspectratio]{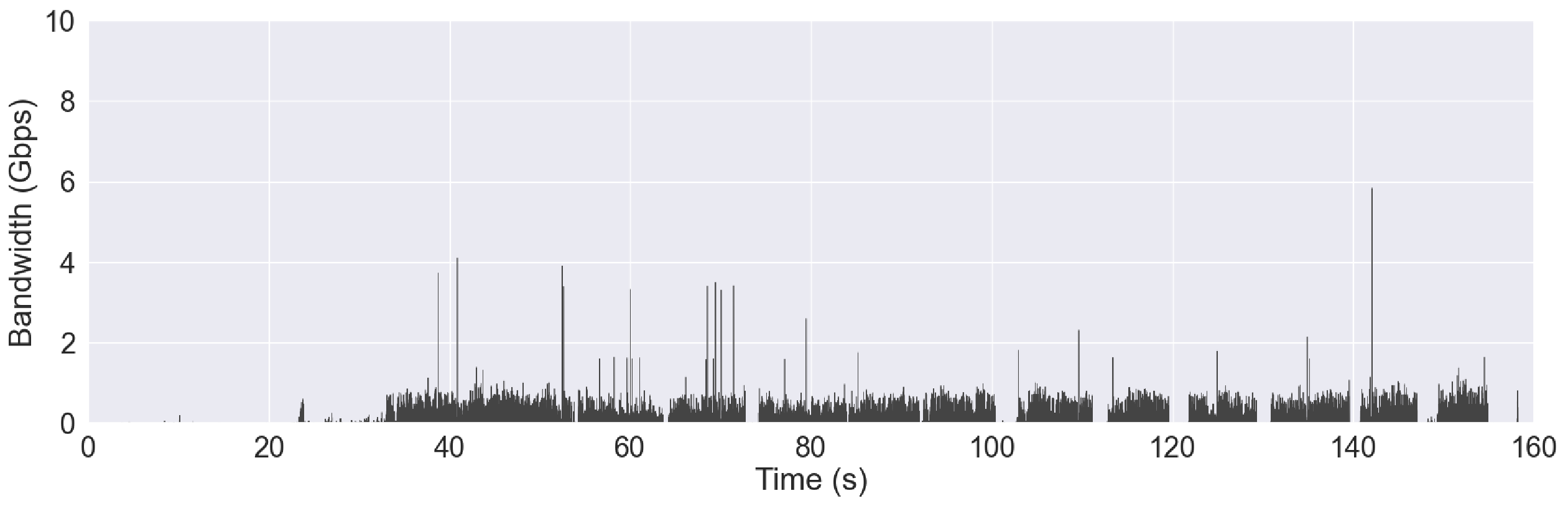} & 
            \includegraphics[width=0.5\columnwidth, keepaspectratio]{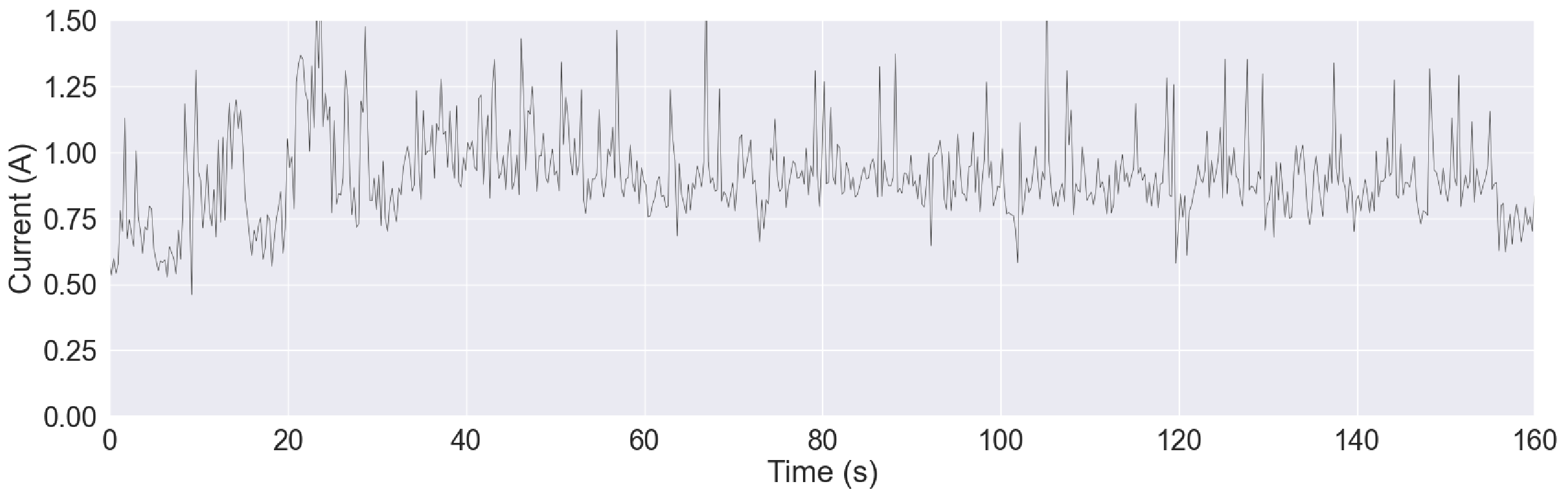} \\    
            {I. Bandwidth, 4K resolution, 4G } &
            {J. Current, 4K resolution, 4G}\\
            \includegraphics[width=0.5\columnwidth, keepaspectratio]{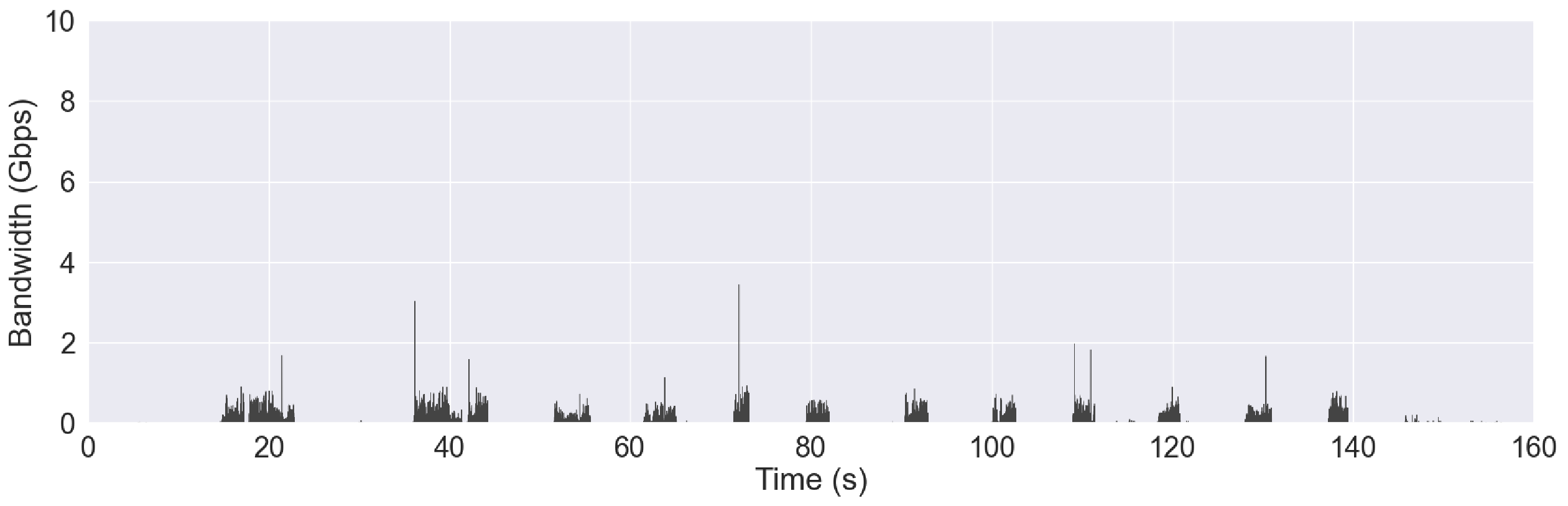} &
            \includegraphics[width=0.5\columnwidth, keepaspectratio]{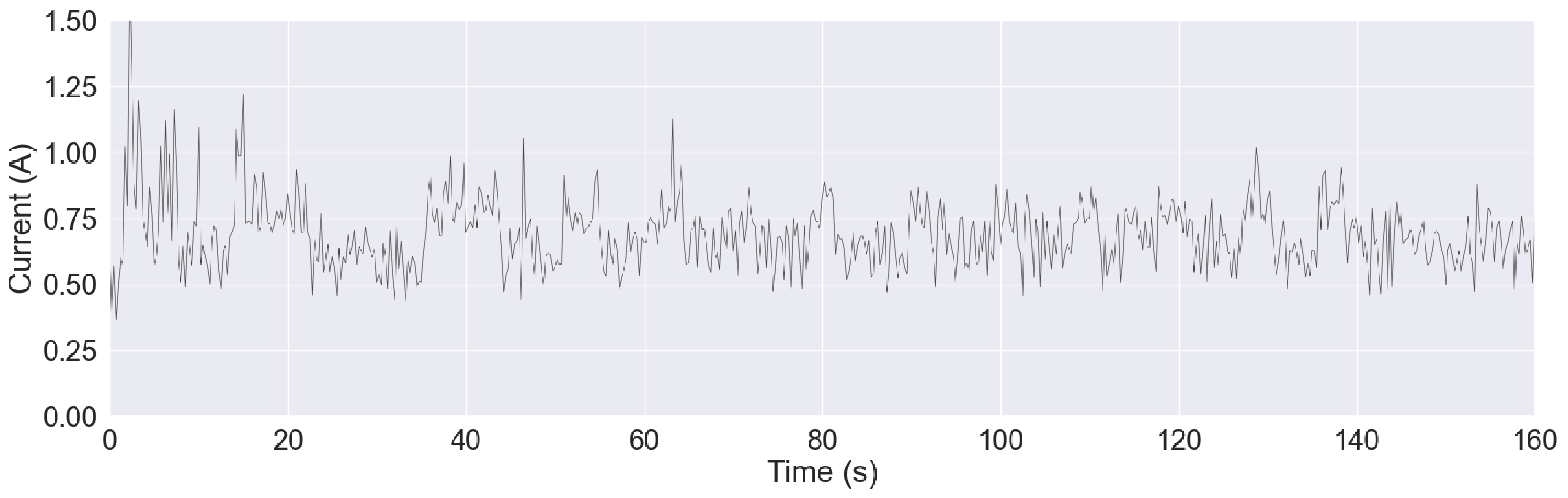}\\
            {K. Bandwidth, 1080p resolution, 4G} &
            {L. Current, 1080p resolution, 4G}\\
            \includegraphics[width=0.5\columnwidth, keepaspectratio]{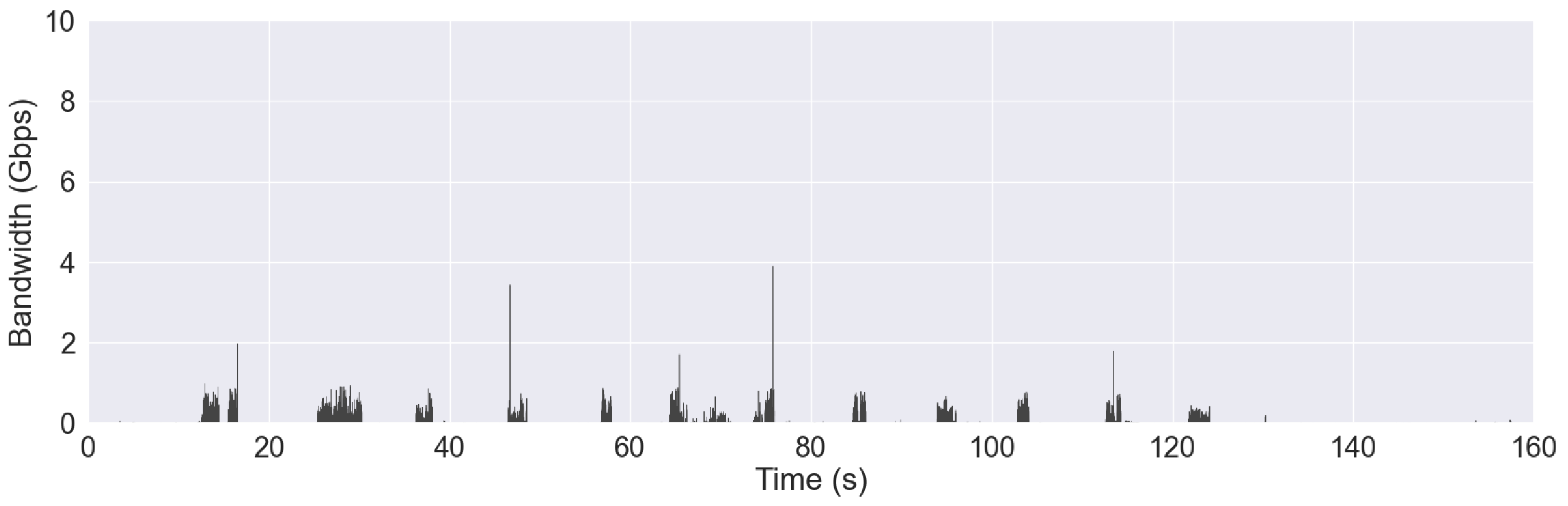} &
            \includegraphics[width=0.5\columnwidth, keepaspectratio]{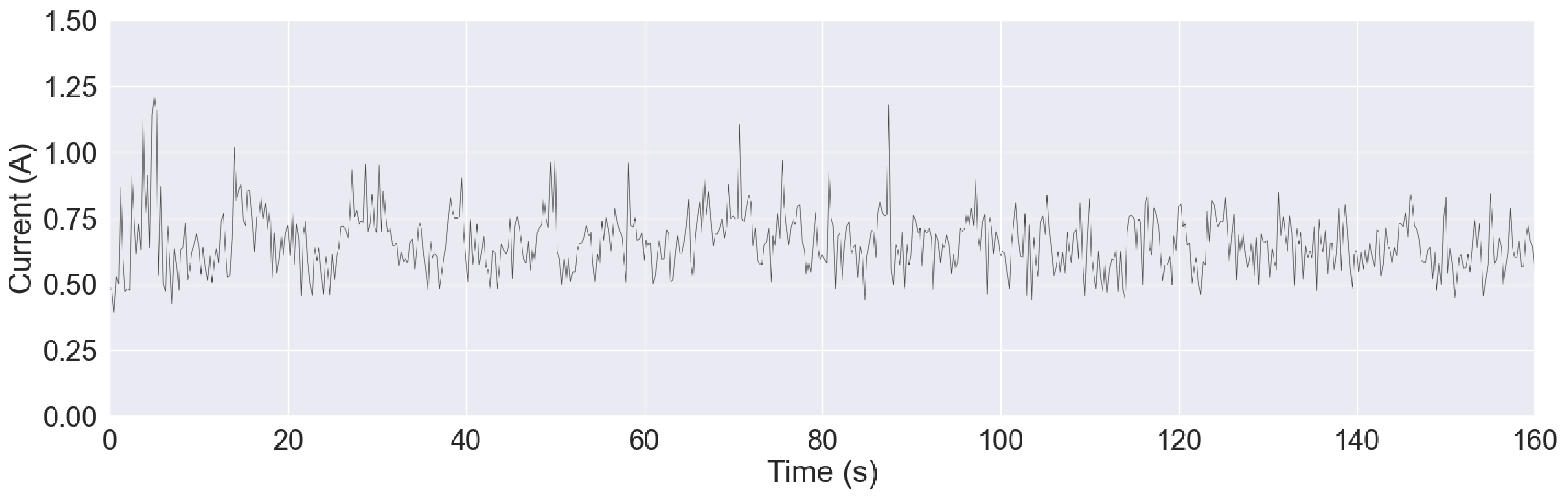} \\
            {M. Bandwidth, 720p resolution, 4G} &
            {N. Current, 720p resolution, 4G}\\
        \end{tabular}
        \caption{Bandwidth and current measurements for \textit{operator 2}.}
     \label{fig:table_of_bandwidths}
\end{figure*}
\twocolumn
\clearpage

\pagebreak

\bibliographystyle{IEEEtran}
\bibliography{refs}

\begin{thebibliography}{10}
\providecommand{\url}[1]{#1}
\csname url@samestyle\endcsname
\providecommand{\newblock}{\relax}
\providecommand{\bibinfo}[2]{#2}
\providecommand{\BIBentrySTDinterwordspacing}{\spaceskip=0pt\relax}
\providecommand{\BIBentryALTinterwordstretchfactor}{4}
\providecommand{\BIBentryALTinterwordspacing}{\spaceskip=\fontdimen2\font plus
\BIBentryALTinterwordstretchfactor\fontdimen3\font minus
  \fontdimen4\font\relax}
\providecommand{\BIBforeignlanguage}[2]{{%
\expandafter\ifx\csname l@#1\endcsname\relax
\typeout{** WARNING: IEEEtran.bst: No hyphenation pattern has been}%
\typeout{** loaded for the language `#1'. Using the pattern for}%
\typeout{** the default language instead.}%
\else
\language=\csname l@#1\endcsname
\fi
#2}}
\providecommand{\BIBdecl}{\relax}
\BIBdecl

\bibitem{iot_devices}
\BIBentryALTinterwordspacing
A.~Holst, ``Number of internet of things (iot) connected devices worldwide from
  2019 to 2030.'' [Online]. Available:
  \url{https://www.statista.com/statistics/1183457/iot-connected-devices-worldwide/}
\BIBentrySTDinterwordspacing

\bibitem{cisco_report}
\BIBentryALTinterwordspacing
Cisco, ``Cisco visual networking index: Global mobile data traffic forecast
  update,'' 2018. [Online]. Available:
  \url{http://www.cisco.com/c/en/us/solutions/collateral/servi
  ce-provider/visual-networking-indexvni/white_paper_c11- 520862.pdf}
\BIBentrySTDinterwordspacing

\bibitem{global_traffic}
\BIBentryALTinterwordspacing
L.~Ceci, ``Mobile internet usage worldwide - statistics i\& facts,'' 2022.
  [Online]. Available:
  \url{https://www.statista.com/topics/779/mobile-internet/#dossierKeyfigures}
\BIBentrySTDinterwordspacing

\bibitem{ibm_report}
\BIBentryALTinterwordspacing
IBM, ``What is edge computing?'' 2020. [Online]. Available:
  \url{https://www.ibm.com/cloud/what-is-edge-computing}
\BIBentrySTDinterwordspacing

\bibitem{8818578}
J.~Hsu, ``How youtube led to google's cloud-gaming service: The tech that made
  youtube work everywhere promises to do the same for games - [news],''
  \emph{IEEE Spectrum}, vol.~56, no.~09, pp. 9--10, 2019.

\bibitem{5g_latency}
\BIBentryALTinterwordspacing
Thales, ``5g technology and networks (speed, use cases, rollout),'' 2022.
  [Online]. Available:
  \url{https://www.thalesgroup.com/en/markets/digital-identity-and-security/mobile/inspired/5G#:~:text=5G%20technology%20offers%20an%20extremely,1%2F1000%20of%20a%20second}
\BIBentrySTDinterwordspacing

\bibitem{youtube_ref}
\BIBentryALTinterwordspacing
J.~Sheperd, ``23 essential youtube statistics you need to know in 2023,'' 2023.
  [Online]. Available:
  \url{https://thesocialshepherd.com/blog/youtube-statistics#:~:text=YouTube%20has%202.1%20billion%20monthly,122%20million%20users%20per%20day.}
\BIBentrySTDinterwordspacing

\bibitem{global_intetnet}
\BIBentryALTinterwordspacing
Ericsson, ``Mobile data traffic outlook,'' 2022. [Online]. Available:
  \url{https://www.ericsson.com/en/reports-and-papers/mobility-report/dataforecasts/mobile-traffic-forecast}
\BIBentrySTDinterwordspacing

\bibitem{10.1145/2742647.2742656}
\BIBentryALTinterwordspacing
K.~Lee, D.~Chu, E.~Cuervo, J.~Kopf, Y.~Degtyarev, S.~Grizan, A.~Wolman, and
  J.~Flinn, ``Outatime: Using speculation to enable low-latency continuous
  interaction for mobile cloud gaming,'' in \emph{Proceedings of MobiSys '15},
  ser. MobiSys '15.\hskip 1em plus 0.5em minus 0.4em\relax New York, NY, USA:
  Association for Computing Machinery, 2015, p. 151–165. [Online]. Available:
  \url{https://doi.org/10.1145/2742647.2742656}
\BIBentrySTDinterwordspacing

\bibitem{edge_copmuting}
\BIBentryALTinterwordspacing
IBM, ``What is edge computing?'' 2022. [Online]. Available:
  \url{https://developer.ibm.com/articles/what-is-edge-computing/}
\BIBentrySTDinterwordspacing

\bibitem{9987496}
S.~Mohebi, F.~Michelinakis, A.~Elmokashfi, O.~Grøndalen, K.~Mahmood, and
  A.~Zanella, ``Sectors, beams and environmental impact on the performance of
  commercial 5g mmwave cells: An empirical study,'' \emph{IEEE Access},
  vol.~10, pp. 133\,309--133\,323, 2022.

\bibitem{electronics9111867}
\BIBentryALTinterwordspacing
D.~Pimienta-del Valle, L.~Mendo, J.~M. Riera, and P.~Garcia-del Pino, ``Indoor
  los propagation measurements and modeling at 26, 32, and 39 ghz
  millimeter-wave frequency bands,'' \emph{Electronics}, vol.~9, no.~11, 2020.
  [Online]. Available: \url{https://www.mdpi.com/2079-9292/9/11/1867}
\BIBentrySTDinterwordspacing

\bibitem{https://doi.org/10.1002/ett.3311}
\BIBentryALTinterwordspacing
M.~Khalily, S.~Taheri, S.~Payami, M.~Ghoraishi, and R.~Tafazolli, ``Indoor
  wideband directional millimeter wave channel measurements and analysis at 26
  ghz, 32 ghz, and 39 ghz,'' \emph{Transactions on Emerging Telecommunications
  Technologies}, vol.~29, no.~10, p. e3311, 2018, e3311 ett.3311. [Online].
  Available: \url{https://onlinelibrary.wiley.com/doi/abs/10.1002/ett.3311}
\BIBentrySTDinterwordspacing

\bibitem{8641429}
Z.~Lin, X.~Du, H.-H. Chen, B.~Ai, Z.~Chen, and D.~Wu, ``Millimeter-wave
  propagation modeling and measurements for 5g mobile networks,'' \emph{IEEE
  Wireless Communications}, vol.~26, no.~1, pp. 72--77, 2019.

\bibitem{7811193}
X.~Zhao, S.~Li, Q.~Wang, M.~Wang, S.~Sun, and W.~Hong, ``Channel measurements,
  modeling, simulation and validation at 32 ghz in outdoor microcells for 5g
  radio systems,'' \emph{IEEE Access}, vol.~5, pp. 1062--1072, 2017.

\bibitem{7954664}
J.~Ko, Y.-J. Cho, S.~Hur, T.~Kim, J.~Park, A.~F. Molisch, K.~Haneda, M.~Peter,
  D.-J. Park, and D.-H. Cho, ``Millimeter-wave channel measurements and
  analysis for statistical spatial channel model in in-building and urban
  environments at 28 ghz,'' \emph{IEEE Transactions on Wireless
  Communications}, vol.~16, no.~9, pp. 5853--5868, 2017.

\bibitem{8419179}
S.~Hur, H.~Yu, J.~Park, W.~Roh, C.~U. Bas, R.~Wang, and A.~F. Molisch,
  ``Feasibility of mobility for millimeter-wave systems based on channel
  measurements,'' \emph{IEEE Communications Magazine}, vol.~56, no.~7, pp.
  56--63, 2018.

\bibitem{8647379}
O.~Al-Saadeh, G.~Wikstrom, J.~Sachs, I.~Thibault, and D.~Lister, ``End-to-end
  latency and reliability performance of 5g in london,'' in \emph{2018 IEEE
  Global Communications Conference (GLOBECOM)}, 2018, pp. 1--6.

\bibitem{10.1145/3387514.3405882}
\BIBentryALTinterwordspacing
D.~Xu, A.~Zhou, X.~Zhang, G.~Wang, X.~Liu, C.~An, Y.~Shi, L.~Liu, and H.~Ma,
  ``Understanding operational 5g: A first measurement study on its coverage,
  performance and energy consumption,'' in \emph{Proceedings of the Annual
  Conference of the ACM Special Interest Group on Data Communication on the
  Applications, Technologies, Architectures, and Protocols for Computer
  Communication}, ser. SIGCOMM '20.\hskip 1em plus 0.5em minus 0.4em\relax New
  York, NY, USA: Association for Computing Machinery, 2020, p. 479–494.
  [Online]. Available: \url{https://doi.org/10.1145/3387514.3405882}
\BIBentrySTDinterwordspacing

\bibitem{10.1145/3366423.3380169}
\BIBentryALTinterwordspacing
A.~Narayanan, E.~Ramadan, J.~Carpenter, Q.~Liu, Y.~Liu, F.~Qian, and Z.-L.
  Zhang, ``A first look at commercial 5g performance on smartphones,'' in
  \emph{Proceedings of The Web Conference 2020}, ser. WWW '20.\hskip 1em plus
  0.5em minus 0.4em\relax New York, NY, USA: Association for Computing
  Machinery, 2020, p. 894–905. [Online]. Available:
  \url{https://doi.org/10.1145/3366423.3380169}
\BIBentrySTDinterwordspacing

\bibitem{10.1145/3452296.3472923}
\BIBentryALTinterwordspacing
A.~Narayanan, X.~Zhang, R.~Zhu, A.~Hassan, S.~Jin, X.~Zhu, X.~Zhang, D.~Rybkin,
  Z.~Yang, Z.~M. Mao, F.~Qian, and Z.-L. Zhang, ``A variegated look at 5g in
  the wild: Performance, power, and qoe implications,'' in \emph{Proceedings of
  the 2021 ACM SIGCOMM 2021 Conference}, ser. SIGCOMM '21.\hskip 1em plus 0.5em
  minus 0.4em\relax New York, NY, USA: Association for Computing Machinery,
  2021, p. 610–625. [Online]. Available:
  \url{https://doi.org/10.1145/3452296.3472923}
\BIBentrySTDinterwordspacing

\bibitem{10.1145/3419394.3423629}
\BIBentryALTinterwordspacing
A.~Narayanan, E.~Ramadan, R.~Mehta, X.~Hu, Q.~Liu, R.~A.~K. Fezeu, U.~K.
  Dayalan, S.~Verma, P.~Ji, T.~Li, F.~Qian, and Z.-L. Zhang, ``Lumos5g: Mapping
  and predicting commercial mmwave 5g throughput,'' in \emph{Proceedings of the
  ACM Internet Measurement Conference}, ser. IMC '20.\hskip 1em plus 0.5em
  minus 0.4em\relax New York, NY, USA: Association for Computing Machinery,
  2020, p. 176–193. [Online]. Available:
  \url{https://doi.org/10.1145/3419394.3423629}
\BIBentrySTDinterwordspacing

\bibitem{youtube_edge}
\BIBentryALTinterwordspacing
scaleyourapp.com, ``Youtube database - how does it store so many videos without
  running out of storgae space?'' [Online]. Available:
  \url{https://scaleyourapp.com/youtube-database-how-does-it-store-so-many-videos-without-running-out-of-storage-space/#:~:text=YouTube%20uses%20low%2Dlatency%2C%20low,it%20from%20the%20origin%20server.}
\BIBentrySTDinterwordspacing

\bibitem{10.1145/1698750.1698753}
\BIBentryALTinterwordspacing
N.~Hopper, E.~Y. Vasserman, and E.~Chan-TIN, ``How much anonymity does network
  latency leak?'' \emph{ACM Trans. Inf. Syst. Secur.}, vol.~13, no.~2, mar
  2010. [Online]. Available: \url{https://doi.org/10.1145/1698750.1698753}
\BIBentrySTDinterwordspacing

\bibitem{6524403}
T.~Gopinath, A.~R. Kumar, and R.~Sharma, ``Performance evaluation of tcp and
  udp over wireless ad-hoc networks with varying traffic loads,'' in \emph{2013
  International Conference on Communication Systems and Network Technologies},
  2013, pp. 281--285.

\bibitem{10.1145/3098822.3098842}
\BIBentryALTinterwordspacing
A.~Langley, A.~Riddoch, A.~Wilk, A.~Vicente, C.~Krasic, D.~Zhang, F.~Yang,
  F.~Kouranov, I.~Swett, J.~Iyengar, J.~Bailey, J.~Dorfman, J.~Roskind,
  J.~Kulik, P.~Westin, R.~Tenneti, R.~Shade, R.~Hamilton, V.~Vasiliev, W.-T.
  Chang, and Z.~Shi, ``The quic transport protocol: Design and internet-scale
  deployment,'' in \emph{Proceedings of the Conference of the ACM Special
  Interest Group on Data Communication}, ser. SIGCOMM '17.\hskip 1em plus 0.5em
  minus 0.4em\relax New York, NY, USA: Association for Computing Machinery,
  2017, p. 183–196. [Online]. Available:
  \url{https://doi.org/10.1145/3098822.3098842}
\BIBentrySTDinterwordspacing

\bibitem{tcpdump}
\BIBentryALTinterwordspacing
T.~T. Group, ``tcpdump,'' 2022. [Online]. Available:
  \url{https://www.tcpdump.org/}
\BIBentrySTDinterwordspacing

\bibitem{5496372}
S.~Wang, D.~Xu, and S.~Yan, ``Analysis and application of wireshark in tcp/ip
  protocol teaching,'' in \emph{2010 International Conference on E-Health
  Networking Digital Ecosystems and Technologies (EDT)}, vol.~2, 2010, pp.
  269--272.

\bibitem{perfetto}
\BIBentryALTinterwordspacing
Perfetto, ``Perfetto - system profiling, app tracing and trace analysis.''
  [Online]. Available: \url{https://perfetto.dev/docs/}
\BIBentrySTDinterwordspacing

\bibitem{android_hal}
\BIBentryALTinterwordspacing
Android, ``Android health.'' [Online]. Available:
  \url{https://source.android.com/devices/tech/health}
\BIBentrySTDinterwordspacing

\bibitem{10.1145/3404868.3406664}
\BIBentryALTinterwordspacing
I.~Livadariu, T.~Dreibholz, A.~S. Al-Selwi, H.~Bryhni, O.~Lysne,
  S.~Bj\o{}rnstad, and A.~Elmokashfi, ``On the accuracy of country-level ip
  geolocation,'' in \emph{Proceedings of the Applied Networking Research
  Workshop}, ser. ANRW '20.\hskip 1em plus 0.5em minus 0.4em\relax New York,
  NY, USA: Association for Computing Machinery, 2020, p. 67–73. [Online].
  Available: \url{https://doi.org/10.1145/3404868.3406664}
\BIBentrySTDinterwordspacing

\bibitem{10.1145/3131365.3131380}
\BIBentryALTinterwordspacing
M.~Gharaibeh, A.~Shah, B.~Huffaker, H.~Zhang, R.~Ensafi, and C.~Papadopoulos,
  ``A look at router geolocation in public and commercial databases,'' in
  \emph{Proceedings of IMC '17}, ser. IMC '17.\hskip 1em plus 0.5em minus
  0.4em\relax New York, NY, USA: Association for Computing Machinery, 2017, p.
  463–469. [Online]. Available: \url{https://doi.org/10.1145/3131365.3131380}
\BIBentrySTDinterwordspacing

\bibitem{10.1007/978-3-540-71617-4_26}
B.~Gueye, S.~Uhlig, and S.~Fdida, ``Investigating the imprecision of ip
  block-based geolocation,'' in \emph{Passive and Active Network Measurement},
  S.~Uhlig, K.~Papagiannaki, and O.~Bonaventure, Eds.\hskip 1em plus 0.5em
  minus 0.4em\relax Berlin, Heidelberg: Springer Berlin Heidelberg, 2007, pp.
  237--240.

\bibitem{10.1145/1971162.1971171}
\BIBentryALTinterwordspacing
I.~Poese, S.~Uhlig, M.~A. Kaafar, B.~Donnet, and B.~Gueye, ``Ip geolocation
  databases: Unreliable?'' \emph{SIGCOMM Comput. Commun. Rev.}, vol.~41, no.~2,
  p. 53–56, apr 2011. [Online]. Available:
  \url{https://doi.org/10.1145/1971162.1971171}
\BIBentrySTDinterwordspacing

\bibitem{band_spectrum}
\BIBentryALTinterwordspacing
Nokia, ``5g spectrum bands explained - low, mid and high bands,'' 2020.
  [Online]. Available:
  \url{https://www.nokia.com/thought-leadership/articles/spectrum-bands-5g-world/}
\BIBentrySTDinterwordspacing

\bibitem{latency_spectrum}
\BIBentryALTinterwordspacing
B.~Ronan~McLaughlin, ``5g low latency requirements,'' 2023. [Online].
  Available: \url{https://broadbandlibrary.com/5g-low-latency-requirements/}
\BIBentrySTDinterwordspacing

\bibitem{s22010026}
\BIBentryALTinterwordspacing
R.~Dangi, P.~Lalwani, G.~Choudhary, I.~You, and G.~Pau, ``Study and
  investigation on 5g technology: A systematic review,'' \emph{Sensors},
  vol.~22, no.~1, 2022. [Online]. Available:
  \url{https://www.mdpi.com/1424-8220/22/1/26}
\BIBentrySTDinterwordspacing

\bibitem{malviya_panigrahi_kartikeyan_2017}
L.~Malviya, R.~K. Panigrahi, and M.~V. Kartikeyan, ``Mimo antennas with
  diversity and mutual coupling reduction techniques: a review,''
  \emph{International Journal of Microwave and Wireless Technologies}, vol.~9,
  no.~8, p. 1763–1780, 2017.

\bibitem{band_spectrum2}
\BIBentryALTinterwordspacing
Nybsys, ``Low to high 5g banhds explained,'' 2023. [Online]. Available:
  \url{https://nybsys.com/5g-bands/#:~:text=The%20data%20speed%20is%20typically,reach%20up%20to%202%20Gbps.&text=The%20mid%20bands%20typically%20use%20a%20channel%20bandwidth%20of%20100MHz.&text=Mid%2Dband%205G%20has%20a,around%2010%20milliseconds%20or%20less.}
\BIBentrySTDinterwordspacing

\bibitem{7469313}
B.~Farhang-Boroujeny and H.~Moradi, ``Ofdm inspired waveforms for 5g,''
  \emph{IEEE Communications Surveys I\& Tutorials}, vol.~18, no.~4, pp.
  2474--2492, 2016.

\bibitem{bandwidth_maximum}
\BIBentryALTinterwordspacing
Thales, ``5g technology and networks (speed, use cases, rollout),'' 2023.
  [Online]. Available:
  \url{https://www.thalesgroup.com/en/markets/digital-identity-and-security/mobile/inspired/5G#:~:text=5G%20technology%20offers%20an%20extremely,1%2F1000%20of%20a%20second}
\BIBentrySTDinterwordspacing

\bibitem{7096297}
G.~Wu, C.~Yang, S.~Li, and G.~Y. Li, ``Recent advances in energy-efficient
  networks and their application in 5g systems,'' \emph{IEEE Wireless
  Communications}, vol.~22, no.~2, pp. 145--151, 2015.

\bibitem{nvidia_cloud_gaming}
\BIBentryALTinterwordspacing
Nvidia, ``Geforce now. system requirements,'' 2020. [Online]. Available:
  \url{https://www.nvidia.com/en-us/geforce-now/system-reqs/#:~:text=GeForce%20NOW%20requires%20at%20least,how%20to%20test%20your%20network.}
\BIBentrySTDinterwordspacing

\bibitem{autonomous_driving}
\BIBentryALTinterwordspacing
DriveU.auto, ``Using 5g for teleoperation of autonomous vehicles,'' 2022.
  [Online]. Available:
  \url{https://driveu.auto/blog/using-5g-for-teleoperation-of-autonomous-vehicles/#:~:text=5G%20networks%20and%20teleoperation%20requirements&text=URLLC%20(Ultra%20Reliable%20Low%20Latency,1%20millisecond%20for%20data%20transmission.}
\BIBentrySTDinterwordspacing

\bibitem{ar_vr_edge}
\BIBentryALTinterwordspacing
M.~Bamforth, ``5g and ar/vr: Transformative use cases with edge computing,''
  2022. [Online]. Available:
  \url{https://stlpartners.com/articles/edge-computing/5g-edge-ar-vr-use-cases/#:~:text=AR%2FVR%20latency%20requirements%20demand%20edge&text=End%2Dto%2Dend%20latency%20should,shown%20on%20the%20diagram%20below.}
\BIBentrySTDinterwordspacing

\bibitem{8337920}
X.~Lin, V.~Yajnanarayana, S.~D. Muruganathan, S.~Gao, H.~Asplund, H.-L.
  Maattanen, M.~Bergstrom, S.~Euler, and Y.-P.~E. Wang, ``The sky is not the
  limit: Lte for unmanned aerial vehicles,'' \emph{IEEE Communications
  Magazine}, vol.~56, no.~4, pp. 204--210, 2018.

\bibitem{amazon_air_prime}
\BIBentryALTinterwordspacing
Amazon, ``Amazon prime air prepares for drone deliveries,'' 2022. [Online].
  Available:
  \url{https://www.aboutamazon.com/news/transportation/amazon-prime-air-prepares-for-drone-deliveries}
\BIBentrySTDinterwordspacing

\bibitem{8434268}
H.~Ji, S.~Park, and B.~Shim, ``Sparse vector coding for ultra reliable and low
  latency communications,'' \emph{IEEE Transactions on Wireless
  Communications}, vol.~17, no.~10, pp. 6693--6706, 2018.

\bibitem{xbox_cloud_gaming}
\BIBentryALTinterwordspacing
Xbox, ``Xbox cloud gaming (beta),'' 2020. [Online]. Available:
  \url{https://www.xbox.com/en-GB/cloud-gaming}
\BIBentrySTDinterwordspacing

\bibitem{10.5555/3485849.3485850}
J.~Bulman and P.~Garraghan, ``A cloud gaming framework for dynamic graphical
  rendering towards achieving distributed game engines,'' in \emph{Proceedings
  of the 12th USENIX Conference on Hot Topics in Cloud Computing}, ser.
  HotCloud'20.\hskip 1em plus 0.5em minus 0.4em\relax USA: USENIX Association,
  2020.

\bibitem{9605447}
M.~Lujan, M.~McCrary, B.~W. Ford, and Z.~Zong, ``Vulkan vs opengl es:
  Performance and energy efficiency comparison on the big.little
  architecture,'' in \emph{2021 IEEE International Conference on Networking,
  Architecture and Storage (NAS)}, 2021, pp. 1--8.

\bibitem{number_smartphone_users}
\BIBentryALTinterwordspacing
Statista, ``Number of smartphone mobile network subscriptions worldwide from
  2016 to 2022, with forecasts from 2023 to 2028,'' 2022. [Online]. Available:
  \url{https://www.statista.com/statistics/330695/number-of-smartphone-users-worldwide/}
\BIBentrySTDinterwordspacing

\bibitem{wathing_youtube_number}
\BIBentryALTinterwordspacing
------, ``Monthly time spent on the youtube mobile app per user in selected
  markets worldwide in 2022,'' 2022. [Online]. Available:
  \url{https://www.statista.com/statistics/1287283/time-spent-youtube-app-selected-countries/}
\BIBentrySTDinterwordspacing

\bibitem{epa}
\BIBentryALTinterwordspacing
EPA, ``Greenhouse gas equivalencies calculator,'' 2023. [Online]. Available:
  \url{https://www.epa.gov/energy/greenhouse-gas-equivalencies-calculator}
\BIBentrySTDinterwordspacing

\bibitem{reactor_energy}
\BIBentryALTinterwordspacing
AGI, ``How much electricity does a typical nuclear power plant generate?''
  2022. [Online]. Available:
  \url{https://www.americangeosciences.org/critical-issues/faq/how-much-electricity-does-typical-nuclear-power-plant-generate}
\BIBentrySTDinterwordspacing

\bibitem{netflix_energy}
\BIBentryALTinterwordspacing
G.~Kamiya, ``The carbon footprint of streaming video: fact-checking the
  headlines,'' 2020. [Online]. Available:
  \url{https://www.iea.org/commentaries/the-carbon-footprint-of-streaming-video-fact-checking-the-headlines}
\BIBentrySTDinterwordspacing

\bibitem{Xu2014DeterminationOT}
\BIBentryALTinterwordspacing
S.~Xu, M.~Perez, K.~Yang, C.~Perrenot, J.~Felblinger, and J.~Hubert,
  ``Determination of the latency effects on surgical performance and the
  acceptable latency levels in telesurgery using the
  dv-trainer{\textregistered} simulator,'' \emph{Surgical Endoscopy}, vol.~28,
  pp. 2569--2576, 2014. [Online]. Available:
  \url{https://api.semanticscholar.org/CorpusID:32583140}
\BIBentrySTDinterwordspacing

\bibitem{Ebihara2022TeleassessmentOB}
\BIBentryALTinterwordspacing
Y.~Ebihara, E.~Oki, S.~Hirano, H.~Takano, M.~Ota, H.~Morohashi, K.~Hakamada,
  S.~Urushidani, and M.~Mori, ``Tele-assessment of bandwidth limitation for
  remote robotics surgery,'' \emph{Surgery Today}, vol.~52, pp. 1653 -- 1659,
  2022. [Online]. Available:
  \url{https://api.semanticscholar.org/CorpusID:248701591}
\BIBentrySTDinterwordspacing

\end{thebibliography}

\end{document}